# Co-designing with neurodiverse children: Reflections on the experiences and ideas of children with developmental language disorder when creating a digital word-learning intervention

**Rafiah Ansari**[1,2,*]; **Rina R. Wehbe**[3]; **Lizbeth Olivia Escobedo**[3]

[1] Discipline of Speech Pathology, Institute of Health and Wellbeing, Federation University Australia, Berwick, Victoria, Australia

[2] Department of Language and Communication Science, School of Health & Psychological Sciences, City St George's, University of London, London, United Kingdom

[3] Faculty of Computer Science, Dalhousie University, Halifax, Nova Scotia, Canada

**ORCID iDs:** Rafiah Ansari 0000-0003-4163-1100; Rina R. Wehbe 0000-0003-3677-5185; Lizbeth Olivia Escobedo 0000-0003-2783-047X

*** Corresponding author:** Rafiah Ansari, rafiah.ansari@federation.edu.au

## Abstract

Introduction: Developmental language disorder (DLD) is a neurodevelopmental condition often characterised by word-learning difficulties that can lead to significant social and academic challenges. The disorder shares some features with other neurodevelopmental conditions such as autism spectrum disorder (ASD).

Despite affecting 7% of children, the condition has received little coverage in participatory design research. This paper addresses this by reporting on the emotional responses and design outputs of children with DLD following participatory sessions to inform a word-learning intervention.

Method: Principles from learner-centred, cooperative, and accessible co-design approaches were integrated to tailor activities for four children with DLD. Design sessions were refined through ongoing monitoring of the children's experiences.
The data that informed the findings included design artefacts such as children's drawings, structured feedback obtained using sentence-starter prompts, and researcher field notes. Reflexive thematic analysis was applied to the data, complemented by findings from an emotion scale completed before and after sessions.

Results: Children responded positively to design activities that incorporated familiar story characters and videogames, and that allowed a direct contribution of design ideas. Activities with higher cognitive demand led to reduced interactions.
Key design requirements for the word-learning intervention included multimodal, personalised learning opportunities and the use of progress markers. Aligning digital and real-world learning, such as providing in-person adult support, was also important.

Conclusions: Through careful planning and refinement, a bespoke participatory design approach proved meaningful and rewarding for children with DLD, whilst also yielding valuable insights for the development of the intervention.

# 1 Introduction

## 1.1 Background

Developmental language disorder (DLD) is a neurodevelopmental condition commonly identified during the primary-school years (ages 5-11) that can result in significant challenges in everyday communication [1]. It is estimated to affect one in seven children [2] and shares some features with other neurodevelopmental conditions such as autism spectrum disorder (ASD) [1].

Poor word learning can affect up to half of children with DLD [3] and is associated with concerns around academic attainment and social well-being [4-5]. Word-learning difficulties occur because of issues coordinating social, physical, cognitive and/or linguistic signals from different environments to help match sounds to meaning [6]. For example, a child with DLD who is shown a 'microscope' and a 'magnifying glass' for the first time may struggle to differentiate between them when told one is used in laboratories. This is because the child must deduce each instrument's purpose by combining their own understanding with the information being provided.

Interventions for children with DLD aim to directly teach sound and meaning associations during word-learning such as through modelling and repetition, while also providing indirect contextual cues such as demonstrating the word in different conversations at different times [7]. Interventions are usually led by Speech and Language Therapists (SLTs) working collaboratively with teachers to deliver evidence-based therapy within school settings. Language goals depend on individual needs and context but should span the child's interest, curriculum focus, and daily routine [8].

## 1.2 The role of digital technology

The work in this paper hypothesises that a user-friendly digital vocabulary intervention would enable children with DLD to create and coordinate multiple sound and meaning word-learning signals i.e., through text, images, audio and video, as is required for optimum language outcomes.

Existing research provides some indications of the potential of digital vocabulary interventions for children with DLD. One study [9] found greater gains when word-learning content was introduced as videos on mobile devices as well as through books compared to books alone. A limitation was that the children played a passive role as content was pre-selected by the researchers. Another study [10], reported greater gains when a custom-made digital aid prompted the sound of initial letters of target words alongside strategies introduced by a SLT, compared to SLT input alone. Whilst promising, the findings related to the use of audio cues only.

An important consideration for the successful adoption and implementation of a digital interventions is the role of the professional. Guidance for SLT practice advocate a blended-learning approach, where in-person professional support is supplemented rather than replaced by computer-mediated content and delivery [11]. The field of Human-Computer Interaction (HCI) design which is a branch of computer science that associates technology with social and behavioral disciplines [12], is well positioned to guide a digital vocabulary intervention that sits within a blended-learning model.

## 1.3 Involving children in intervention development

A systematic review of digital interventions for children with communication difficulties reported a lack of studies considering the child's intervention preferences [13]. This can be addressed through participatory design research, a HCI approach that directly involves users in the design process to ensure that digital products meet their everyday needs and preferences [14].

Whilst there are no available examples of participatory design to inform interventions for children with DLD, there can be learning from co-design literature that covers neurodevelopmental disorders more broadly. When children's perspectives are considered, communication interventions have been evaluated as acceptable, suitable and engaging by SLTs, teachers, and children with a range of conditions including ASD, cerebral palsy, learning difficulties, and visual impairments [15-17]. Children's involvement tends to focus on sharing ideas to make interventions motivating and engaging, with the rationale that children with learning needs enjoy these contributions but find providing feedback on therapeutic content challenging [18-22].

Using participatory design to enable children with DLD to inform a word-learning intervention requires careful planning in order to understand the child's learning environment and align the participatory design to each child's individual needs. These complexities can be addressed by combining design principles that are learner-centred, collaborative, and accessible. Existing design approaches that can help achieve this are Informant Design [20], Cooperative Inquiry [21] and Diversity for Design [22] as explored further below.

**A learning-centred design approach**

Informant Design [20] involves children in the design of learning systems by exploring not only their views but also their experiences and needs as learners. This includes consideration of the role of support staff and specialist professionals and the influence of the physical learning environment.

An example of use for intervention design comes from Malinverni et al. [23] who created a virtual reality game to develop social skills in children with ASD. The design strategy split the informant design into two consequent stages, the first was collaboration with professional experts [psychologists and psychiatrists] to define learning goals. The second stage used participatory design with children as informants to understand their play and social preferences and transform defined learning goals into an enjoyable game experience.

**A collaborative design approach**

Cooperative Inquiry [21] considers children as design collaborators who can generate novel design ideas emerging from their natural environment and based on their own technological experiences and preferences. Child-friendly collaborative design methods are advocated such as familiar stories to provide design purpose from which children can collaboratively generate and explore novel design suggestions that may otherwise feel abstract and decontextualised [24].

Wilson et al. [25] used craft materials with minimally-verbal children with ASD, supported by their SLT, to elicit play preferences which informed a tangible digitally-enhanced ball to self-express during play. In another example, Millen et al. [26] asked children with ASD, supported by their specialist teacher, what they liked and disliked about computer games they already used to understand design preferences for a virtual, player-collaboration social-skills game.

**An accessible design approach**

Diversity for Design [22] promotes accessible design by focusing on the strengths of neurodiverse children, while supporting their needs. The term neurodiverse is adopted to encourage thinking towards alternative rather than deficit cognitive processing. The framework serves to understand evidence-based patterns of behaviours that characterise a condition, recognise individual diversity, align the environment to preferred learning styles, and provide social support. Continuous cycles of reviewing and responding to the experiences of the children is vital for successful implementation of the accessibility strategies.

Benton et al. [22] applied the framework when working with children who had dyslexia to generate navigation and reward elements for a language and literacy game. Typical profiles of children with dyslexia influenced support strategies such as visual templates for collecting feedback, and sessions in quiet and familiar environments with consistent scheduling. Specialist teachers familiar with the children then advised on best methods for individualising the support.

**Combining design approaches that are accessible, collaborative and learner-centred**

Participatory design with neurodiverse children requires a balance as summarised by Frauenberger et al. [27] in their article on creating technologically-enhanced learning environments: '*When designing with children with disabilities, the issue of beneficence – ensuring that the risk or demand on them does not outweigh the benefits of inclusion – is paramount'* [27, p26].

Some studies have sought to address this by coordinating multiple principles and techniques to suit both the needs of the design and the needs of the children supporting the design. For example, Frauenberger et al. [28] and Makhaeva et al [29] collectively report on a bespoke approach to designing communication support systems for children with ASD which tailored the process to the children and the context. Methods included the Diversity for Design framework [22] to provide an evidence-based foundation for considering the children's strengths, needs, and support mechanisms, which evolved as the children became familiar with the design process and their profile of presentation changed. In parallel, the Cooperative Inquiry approach [21] provided initial design techniques, which were extended to include other methods and design principles depending on the children's individual capacities and preferences, including shifting some of the children's involvement from designing to informing. Further relevant design requirements were obtained from other sources such as specialist teachers who were familiar with the children.

This multi-method approach to creating bespoke participatory design experiences has yet to be applied to children with DLD. This paper therefore aims to:

1. Understand the appropriateness of an accessible, collaborative, and learner-centred design approach for children with DLD by investigating their experience of the process.
2. Understand requirements generated from the design sessions in order to make a digital word-learning intervention engaging and motivating for children with DLD.

## 2 Method

This paper reports on the experiences and outputs following participatory design sessions with children who have DLD to inform a word-learning intervention. The intervention is intended to be used as part of a blended-learning model [11] where the technology augments existing in-person support for children provided by SLTs working with school educators. As advocated in HCI literature [20], the study focused on generating children's design ideas to inform motivational and engagement aspects of the word-learning intervention. The therapeutic content of the intervention is guided by word-learning theory, research, and clinical practice and the reader is directed to Ansari et al. [30] for more detail.

## 2.1 Target population and sample size

The participant eligibility criteria comprised of children with DLD aged 7-11-years (based on target age for the intervention, see 30), with sufficient language competency to reflect and feedback during participatory design sessions (as judged by the children's school and/or SLT). The study aimed to recruit 4 children based on the average participant numbers for participatory design studies involving neurodiverse children as identified in systematic reviews [31-33].

## 2.2 Ethics

Study approval was obtained from the School of Health & Psychological Sciences Research Ethics Committee, City, University of London, UK [ETH2324-1567] and the Health Sciences Research Ethics Board, Dalhousie University, Nova Scotia, Canada [REB # 2023-6880].

## 2.3 Selection and participation of children

The study used convenience sampling to recruit four children who met the criteria for DLD [1]. All the children attended a specialist language unit within a mainstream school and were selected as being eligible by the school's Inclusion Manager, a specialist teacher who actively taught the children and managed school-based additional needs support across the school, with consensus from the school SLT. The school was based in South-West London in a region of higher than national ethnic diversity [34] and a greater proportion of social housing compared to neighbouring areas [35].

Table 1 provides a summary of participant characteristics. It is worth noting that all the children were male, this homogeneity was not intentional but reflects that prevalence of DLD is significantly higher in males than females, with a ratio of 2:1 [36].

To reflect best practice in terms of collaborative support for children with DLD, design sessions were planned in conjunction with the school SLT and Inclusion Manager, with involvement of the Teaching Assistant (TA) who provided day-to-day support under the guidance of the SLT and Inclusion Manager [37]. In addition, and again in keeping with the best practice version of standard practice, direct implementation of the design sessions was supported by the TA and school SLT.

A detailed participant information sheet and an easy-read summary was created for parents to share with their children in advance of the study with the opportunity to ask questions before written parental consent was gained.

Steps taken at the start and throughout the study to ensure the children were happy and comfortable with participating included careful observations, liaising with the support staff attending the sessions (School SLT and TA, both familiar with the children), asking the children directly, and reminding the children that they could stop for any part of the study at any point without giving reason. Care was taken, through discussion with school and parents, to arrange sessions at times that did not conflict with compulsory lessons or those of high interest for the children.

Table 1 Participant demographics

| | **Age** | **Diagnosis** | **Demographics*** |
|---|---|---|---|
| Child 1 (Male) | 11 | DLD | Only English spoken<br>Not on subsidised school meals. |
| Child 2 (Male) | 11 | DLD, speech disorder, ADHD | Only English spoken<br>Subsidised school meals. |
| Child 3 (Male) | 10 | DLD, social skills difficulties | Bilingual [English & French]<br>Not on subsidised school meals. |
| Child 4 (Male) | 10 | DLD, ADHD | Only English spoken<br>Not on subsidised school meals. |

*Demographics refer to whether English is an additional language and socioeconomic status (considered low if in receipt of subsidised school meals)

## 2.4 Design approaches and activities

**Structure and format of design sessions**

In the first instance, the Diversity for Design framework [22] helped generate a standard template for providing an accessible structure and format to design sessions (Table 6.2). This involved collating typical strengths, preferences and difficulties experienced by children with DLD from existing literature, most notably work to define DLD that is based on a multinational and multidisciplinary consensus [1, 38]. Contributors to the consensus included SLTs, educational psychologists, paediatricians and specialist teachers.

Recommended support strategies which aligned to the characteristic profile of children with DLD [39-40] were organised according to the framework components: structuring the environment, adult scaffolding, and tailoring to the individual. The accessibility strategies were then individualised to each child's unique presentation using information collected from the children before the testing session via an 'About Me' profile, which were completed by the children with support from the school SLT, teaching staff, and parents/carers. The profile was created specifically for the study and covered the children's likes, dislikes, strengths and needs (Appendix A).

An example of personalised support is that as all of the children's profiles indicated a liking for physical activities, this was incorporated into the sessions, e.g. bean bag throwing to indicate a preference. Learning a new skill was a common difficulty and was kept to a minimum by using familiar activities where possible. Non-verbal communication opportunities such as pictures and videos were balanced with options for reading, writing, speaking and listening to spoken information. This was done as half of the children identified areas of language and literacy as strengths in their profiles despite diagnoses of language difficulties. This demonstrated the importance of considering the children's perspective in addition to diagnostic profiles when considering support strategies. Regular conversations with the school SLT, Inclusion Manager, and TA before and after each co-design session as well as a mid-study check in with the children's parents helped ensure ongoing relevancy of the support strategies. In addition, the presence of the school SLT and TA within the sessions meant that they were able to provide the individualised support required for each child as sessions progressed.

Table 2 Participatory design framework for children with DLD
Based on the Diversity for Design template [23]

| Presentation | Area of strength | Area of need | Recommended Strategy | | |
|---|---|---|---|---|---|
| | | | **Structuring Environment** | **Adult Scaffolding** | **Tailoring to Individual** |
| Understanding & use of language | | * | Pre-teach vocabulary required for session | Support understanding of information & instructions<br>Support generation & expression of ideas & opinions | Ensure appropriate language content for ability |
| Short-term memory of verbal information. | | * | Ensure consistent session structure<br>End each session with summary & plan for next session | Repeat & reinforce relevant information | |
| Responsiveness to routine & structure | * | | | | |
| Responsive to visual & tactile cuing | * | | Start sessions with visual, written & verbal recap.<br>Use visual timetable with symbols & words to engage & organise activities | | Provide choice of tangible & digital multisensory activities & feedback methods [spoken, drawn, written, recorded audio, video] |
| Responsive to written cuing | * | | | | |
| Responsive to adult social cuing | * | | | Model activities to demonstrate what to do.<br>Encourage engagement | |
| Self-regulation | | * | Provide a distraction free & familiar environment | Support use of strategies to regulate emotionally & physically | Ensure access to objects & activities to support regulation |
| Motivated by achievement | * | | | Support errorless participation | |

## Specific design activities

To plan the content of the design activities a common user-design flow was adopted, moving from understanding and contextualising the user's needs, gaining design ideas, and jointly prototyping (41). The principles of learner-centred, cooperative and accessible design were met by integrating approaches from Informant Design [20], Cooperative Inquiry [21], and Diversity for Design [22]. This was achieved through four iterative design phases: Phase 1: Introducing a story to provide cohesion between sessions; Phase 2: Understanding the user by exploring strengths, needs, and current support; Phase 3: Exploring current systems; Phase 4: Designing a solution. The design phases, corresponding design approaches and specific activities undertaken are all detailed in Table 6.4. Although sessions and activities within each phase were linear, the learning was iterative and often overlapped with previous and later phases. For example, children's design suggestions (Phase 4) gave insight to their needs as a user (Phase 2).

The design activities were customised to children's interests and preferences using information from their 'About Me' profiles (Table 6.3) and through conversations with the Inclusion Manager, TA, and SLT. For example, a superhero theme was recommended by the TA to provide a storytelling structure to the design sessions.

All four children referred positively to their use of technology and gave multiple examples in their 'About Me' profiles. As a result, two videogames were selected to understand children's technology preferences during design sessions: Minecraft 1.20 (Mojang Studios www.minecraft.net), and Dream League Soccer 2024 - DLS24 (First Touch Games Ltd www.ftgames.com). Minecraft is an open-ended video game where users explore, build and share virtual environments. DLS24 is a football simulation game where players build and manage football team. All the children were familiar with both games.

Table **3** Children's individual profiles

| | **Likes** | **Strengths** | **Dislikes** | **Needs** |
|---|---|---|---|---|
| Child 1 (Male) | *'Playing football, penalty shootouts, being with my friends, being on my phone, You Tube, TV, DLS24 [video game]'* | *'Dancing, dribbling in football, making friends, amazing writing'* | *'People butting in'* | *'Sometimes with my work: math, technology, writing, geography'* |
| Child 2 (Male) | *'Cuphead* [video game], *Don't Deal with the Devil* [song linked to Cuphead video game], *DuckTales: Remastered* [video game]' | *'Math, PE, football music, life skills'* | *'Mushrooms'* | *'Things that aren't science'* |
| Child 3 (Male) | *'Sponge Bob* [cartoon], *Minecraft* [video game] *DanTDM, iBallisticSquid, Stampy* [British video gamers & YouTubers], *Netflix, YouTube'* | *'Minecraft, reading, 'speaking in long sentences'* | *'Fortnite, Roblox, modern animation, modern music, maths'* | *'Making friends, understanding the teacher, speaking to new people, focusing on the carpet in class, making stories'* |
| Child 4 (Male) | *'Online Avatars, Minecraft'* | *'Gymnastics, hockey, basketball, football, cycling, art'* | *'Making friends'* | *'Volleyball, talking in front of friends'* |

Table **4.** Summary of participatory design phases, aims and activities

| Phases | Design Approach | Purpose | Activities | |
|---|---|---|---|---|
| | | | Symbol* | Description & key words |
| **Phase 1 Introduce a story for a running theme between design sessions.** | **Cooperative Inquiry** [21, 24]**:** Storytelling to build cohesion between design sessions | **Session 1: DESIGN STORY** Create continuous, meaningful design experiences using an underlying narrative & corresponding story props (e.g. storybooks). Introduce in the first session & continue across sessions. | **Superheroes** | **Superhero theme**: Short introductory session to explain the reason for the design work. Primary author introduced concept of designing a word-learning app to help superheroes with using the right words.<br>Children were then provided with a range of superhero props to engage with & explore. e.g. comic books, masks, pictures. These were integrated into subsequent design sessions.<br>**Key words:** research, co-design, feedback |
| **Phase 2: Understand the user: Explore strengths, needs & current support.** | **Informant Design** [20]: Research on the user's needs as a learner<br>**Diversity for Design** [22]: Design that capitalises on strengths to support needs | **Session 2: WORD GAMES** Understand children's attitude towards words & word-learning activities to help define the system. | Abcd word cloud | **Word Cloud game**: In this session children engaged with a language activity that involved generating high interest words. All participants (children, SLT, TA) shared their favourite superhero words before taking a turn on a familiar group game (Pop Up Pirate).<br>Words were written by child/adult onto a 'cloud' drawn on an A3 wall sheet for all to see.<br>**Key words:** research, co-design, feedback, words |
| | | **Session 3: STRENGTHS & NEEDS** Understand children's perspective of strengths & needs to create a system that supports user's needs through their strengths | **Strengths & Needs** | **Explore Strengths & Needs:** Explored what the words 'strengths' & 'needs' meant to the children. Then asked the children to assign a range of strengths & needs cards [chosen by the support staff] to the superhero characters using a bean bag throwing game. Option was given to generate further characteristics using blank cards.<br>**Key words:** research, co-design, feedback, strengths, needs |
| | | **Session 4: STRATEGY GENERATION** Identify children's preferred word-learning strategies for the system | **Support** | **Think of support activities**: Children asked to identify word learning strategies that superheroes would find helpful.<br>**Key words:** research, co-design, feedback, words, strengths, needs, support |
| | | **Session 5: STRATEGY EVALUATION** Understand children's views of word-learning strategies to inform the system | **Evaluate** | **Evaluate word learning strategies**: Children asked for feedback on different word-learning strategies [like/dislike/unsure/change].<br>**Key words:** research, co-design, feedback, words, strengths, needs, support |
| **Phase 3: Explore current systems** | **Cooperative Inquiry** [21]: 'Technology Immersion' to understand preferences<br>**Diversity for Design** [22]: Sessions themed to user's hobbies & interests | **Session 6: FAVORITE TECH** Understand technology preferences by observing children interacting with existing systems & replicate in proposed system. | **Video Games** | **Share a videogame:** Children provided with familiar videogames that they all enjoyed and asked to describe and share. Videogames were presented on school iPads with permission from the school headteacher.<br>**Key words:** research, co-design, feedback, technology |
| **Phase 4: Design a solution** | **Cooperative Inquiry** [21]: 'Bags of Stuff' to generate ideas | **Session 7: APP DESIGN** Gather children's design suggestions for a word-learning system | **Design an app** | **Design a word learning app for your superhero:** Generated design ideas using key themes from videogame feedback session to structure suggestions. Used scenarios based on superhero characters/games. Provided art & craft materials to inspire creativity.<br>**Key words:** research, co-design, feedback, words, strengths, needs, support, technology |

***Activity symbols created by children's speech & language therapist and introduced to the children as part of a visual timetable.**

## 2.5 Tools

**Emotion scale**

In line with the Diversity for Design approach [22], an emotion scale that was familiar to the children was used to identify and support children's state before and after each design session (Fig 1). The scale is adapted from the Zones of Regulation™ curriculum [39], an evidence-based approach to supporting neurodiverse children with emotion regulation [42-43]. The scale has four colour zones, each representing distinct emotions: Blue represents low alertness and down feeling (sad, tired, unwell, bored); Green: calm state of alertness (calm, happy, focused, ready to learn); Yellow: heightened alertness and elevated emotions (excited, worried, silly, frustrated); Red: extremely high energy and intense, overwhelming feelings (very excited, angry, scared, out of control). Whist the 'Green zone' is considered optimum for learning, other zones may be beneficial depending on the individual child [42].

Children were supported by their SLT and TA to identify their emotional state using the scale or by generating their own description. A judgement was made by the support staff as to whether a child's response indicated an area of need that would impact their participation. The child was then supported using individualised strategies and this was reported in the findings (e.g., one child benefitted from a sensory squeeze toy to help regulate when feeling 'silly').

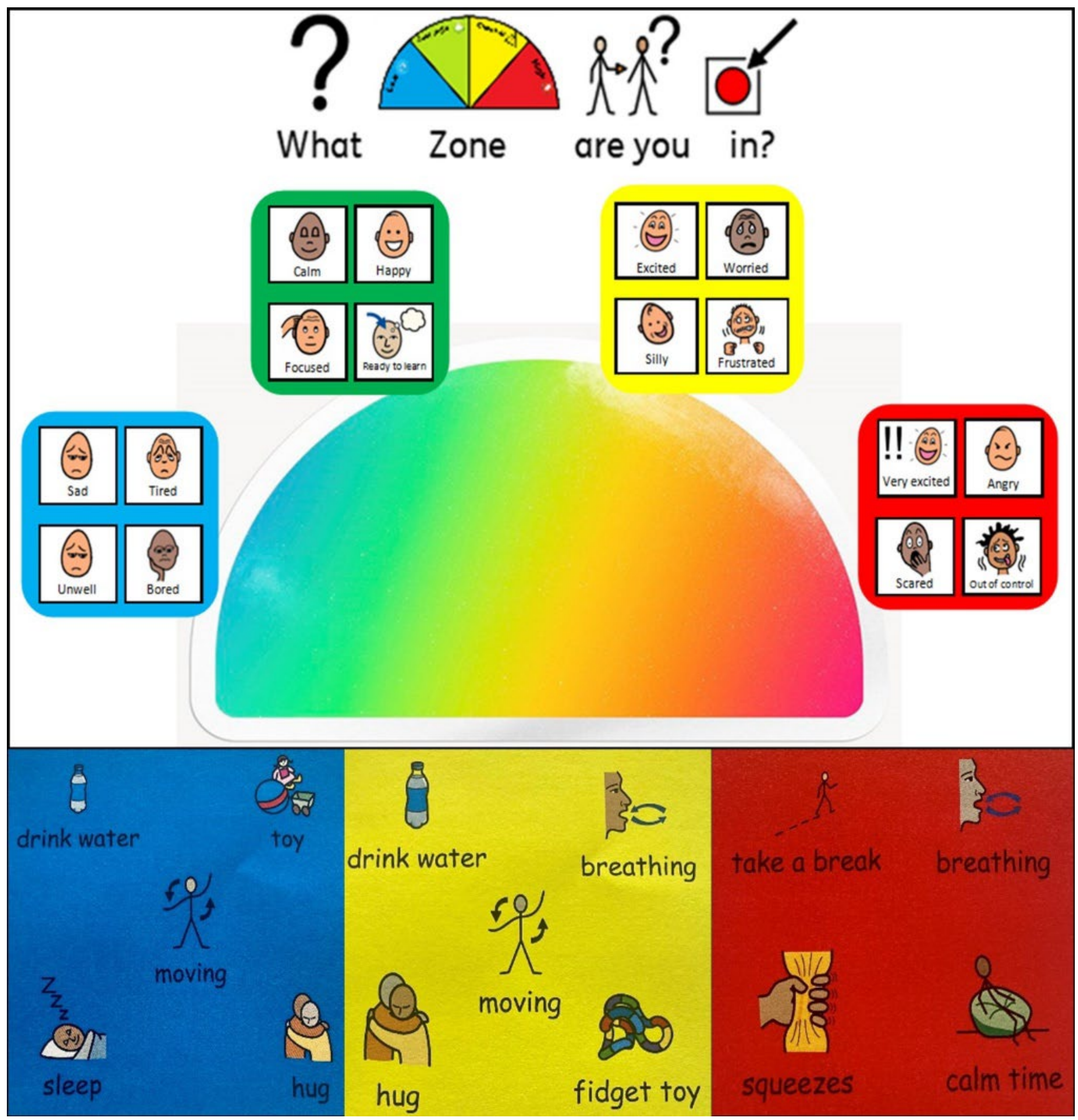

Figure 1 Emotion scale
Used by children as part of their school routine, adapted from The Zones of Regulation™ approach [42] and shared here with permission. The emotion symbols were created using Boardmaker7® and are shared here with permission.

**Sentence prompts for structured feedback**

Structured feedback provided insight into children's thoughts and feelings about the design activities and outputs. This was obtained using sentence starter prompts of '*I like…', 'I'm not sure…', 'I don't like…', 'We can make this better by…*', which is adapted from a method used by Millen et al. [26] when they conducted Cooperative Inquiry design [21] with neurodiverse children. The feedback prompts were presented verbally by the support staff, as well as in writing with a symbol to visually represent the sentences (Fig 2). The children were supported by the support staff to provide feedback using their preferred communication method.

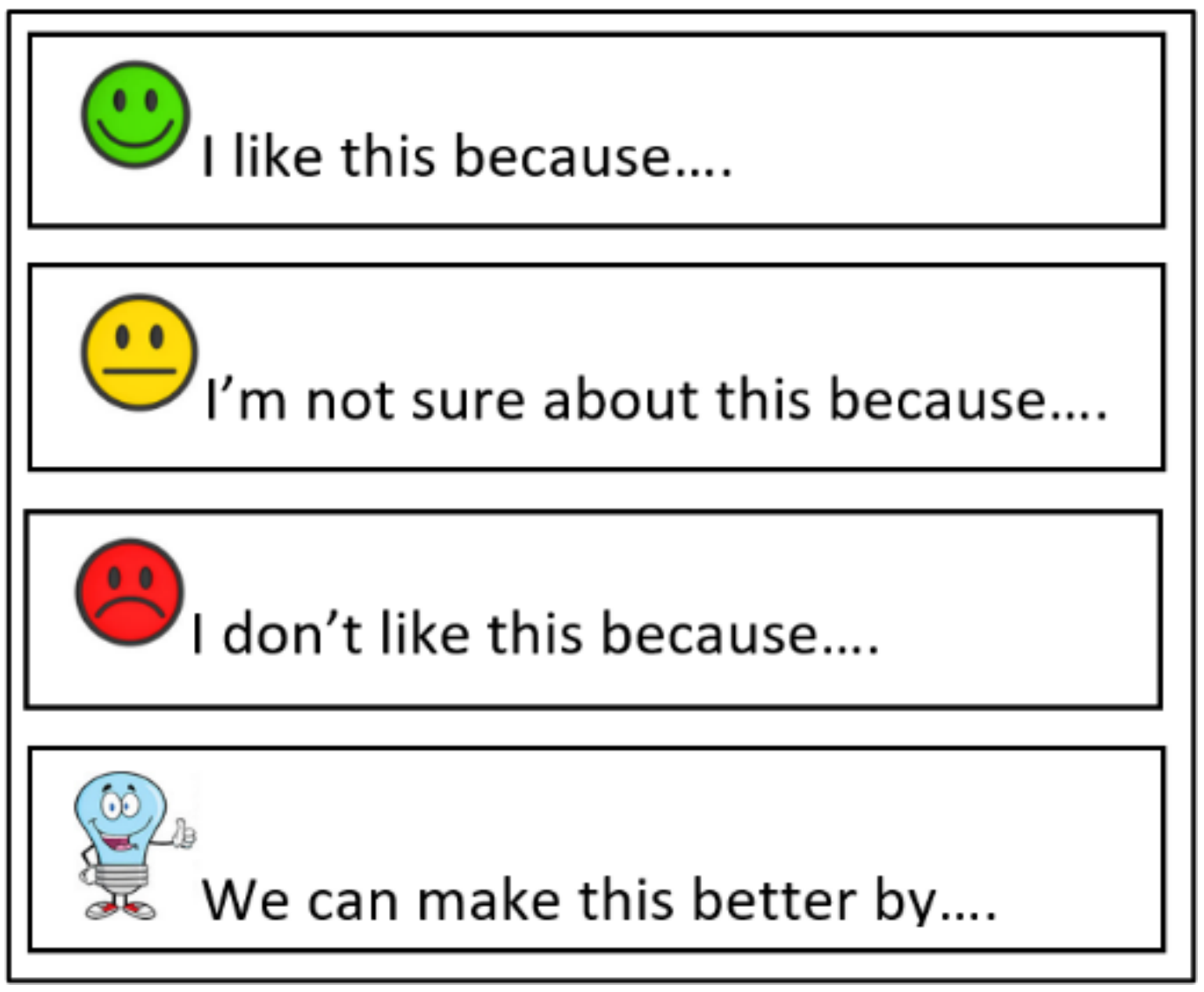

Figure 2 Sentence starter prompts to structure design feedback

## 2.6 Procedure

The study required seven participatory design sessions completed over three months to implement planned design activities. The distribution of the sessions was based on the availability of the children and school staff. The first session lasted 30 minutes and the remaining sessions between 45-60 minutes. All sessions involved the four recruited children with DLD supported by their SLT and TA.

Sessions took place in a quiet room in the children's school to provide a distraction-free, familiar environment. Staff participated in all activities to provide a model for the children to follow and used existing knowledge of the children to provide support such as simplifying language, providing choices, and repeating instructions. The design sessions gave children choice in how they accessed and shared information (e.g., spoken, drawn, written, recorded audio, video).

Each session began with a visual timetable that was created and introduced by the children's SLT to explain planned activities. The first item was always the emotion scale (Fig 1), followed by a recap of the previous session. Key words for the current session were then introduced to ensure shared understanding. This was achieved by writing key words, identified beforehand with support staff, on a wall sheet and asking the children what the words meant to them. Any gaps in knowledge were supported by the children's SLT and TA.

Specific design activities were then completed to inform the word-learning intervention (see Table 3). Each activity was customised to the children's individual interests and preferences and existing staff knowledge. For example, ensuring opportunities for movement breaks as most children referred positively to physical activity in their profiles. After the design activities, children's feedback was gained using sentence prompts (Fig 2) to capture their design experience and areas requiring improvement ahead of subsequent session. Plans for the next session were then shared.

The session ended by repeating the emotion scale (Fig 1) with emotional regulation support provided as needed by the support staff so that there was a smooth transition back to class. Following the final session, children were awarded a £10 Amazon gift voucher and a thank you certificate during a special whole class assembly led by the school's Inclusion Manager.

## 2.7 Data collection and analysis

**Qualitative**

Reflexive thematic analysis [45] was used to investigate children's design experiences and ideas. This approach enabled the generation of key themes from the emerging data allowing a deep understanding of children's diverse perspectives. Three data sources were considered: [i] children's artefacts such as drawings, writings, and video recordings from the design activities; [ii] structured feedback from children following design activities obtained using sentence starter prompts (Fig 2); [iii] field notes made by the first author discretely during sessions and immediately after.

The qualitative analyses began immediately following the first design session with the primary author (RA) examining all generated data. The data was then systematically coded for areas of importance; children's participatory design experiences and children's design requirements. Codes were collated into themes which represented emerging trends, e.g. 'responsiveness to customised learning'. The codes and themes were validated through agreement with the Inclusion Manager, SLT and TA who jointly supported the planning and implementation of design activities until full consensus was reached. Codes from the first design phase were then applied to the next design phase to add depth to the themes. In addition, new codes were identified as more data of interest emerged. These either aligned to existing themes or new themes were generated. Validation was again undertaken through agreement with the staff who supported the design activities. Findings from each phase helped refine subsequent activities as recorded in the results section. This iterative, cumulative analysis continued until all the design phases had been completed and all data of relevance coded, categorised and validated.

### 2.8 Quantitative

To further understand children's design experiences, consideration was given to children's emotion scale responses pre- and post-session (Fig 1). Bar graphs were created to visually represent which emotions were selected and by how many children before and after each design session. This information was triangulated with the qualitative findings.

## 3 Results

### 3.1 What are the experiences of children with DLD when participating in an accessible, collaborative, and learner-centred design approach?

Fig 3 summarises the total count for each emotion before and after individual design sessions and captures how this self-reported measure of children's design experience evolved over the course of the study. This quantitative data was coordinated with the qualitative data from the sessions and presented below according to prominent themes relating to children's design experiences. See Appendix B for ffurther details of the qualitative analysis of children's responses (codebook, themes and subthemes).

**Theme 1: Exploring children's feelings at the start of design sessions indicates factors that may influence children's design experience and readiness for design activities.**

Children's emotional responses at the start of sessions indicated some of the influences of the school environment and routine on their presentation. For example, at the start of the very first session (Introduce Design Story), half of the children (Child 1 and 4) selected '*tired'* for their feelings, with Child 1 stating:

> '*We've just had lunch, so I'm sleepy but I do want to join in.'* (Child 1)

Child 4 agreed with Child 1. Reflections between the primary author (RA) and support staff following the session led to the decision to keep the session time. This was based on three key factors: [i] school confirmed that there was no other time available for the sessions, [ii] none of the children stated they were tired following the session, [iii] and the children's state provided a more realistic picture of how the children may engage with learning in practice.

In session two (Word games), half of the children (Child 2 and 4) selected '*unwell'* to describe their feelings pre-session (one child stated they had a cough and the other a cold). In both cases the TA was aware and in line with the Cooperative Inquiry [21] approach, the children were consulted regarding whether they wished to continue which they did.

Similarly, in session six (Sharing Favourite Tech), Child 2 selected '*sad'* to describe their emotion pre-session. The TA discussed this further with the child in private who informed that the emotion was related to an event in the playground prior to the session but it had been worked through and the child wished to take part in the session.

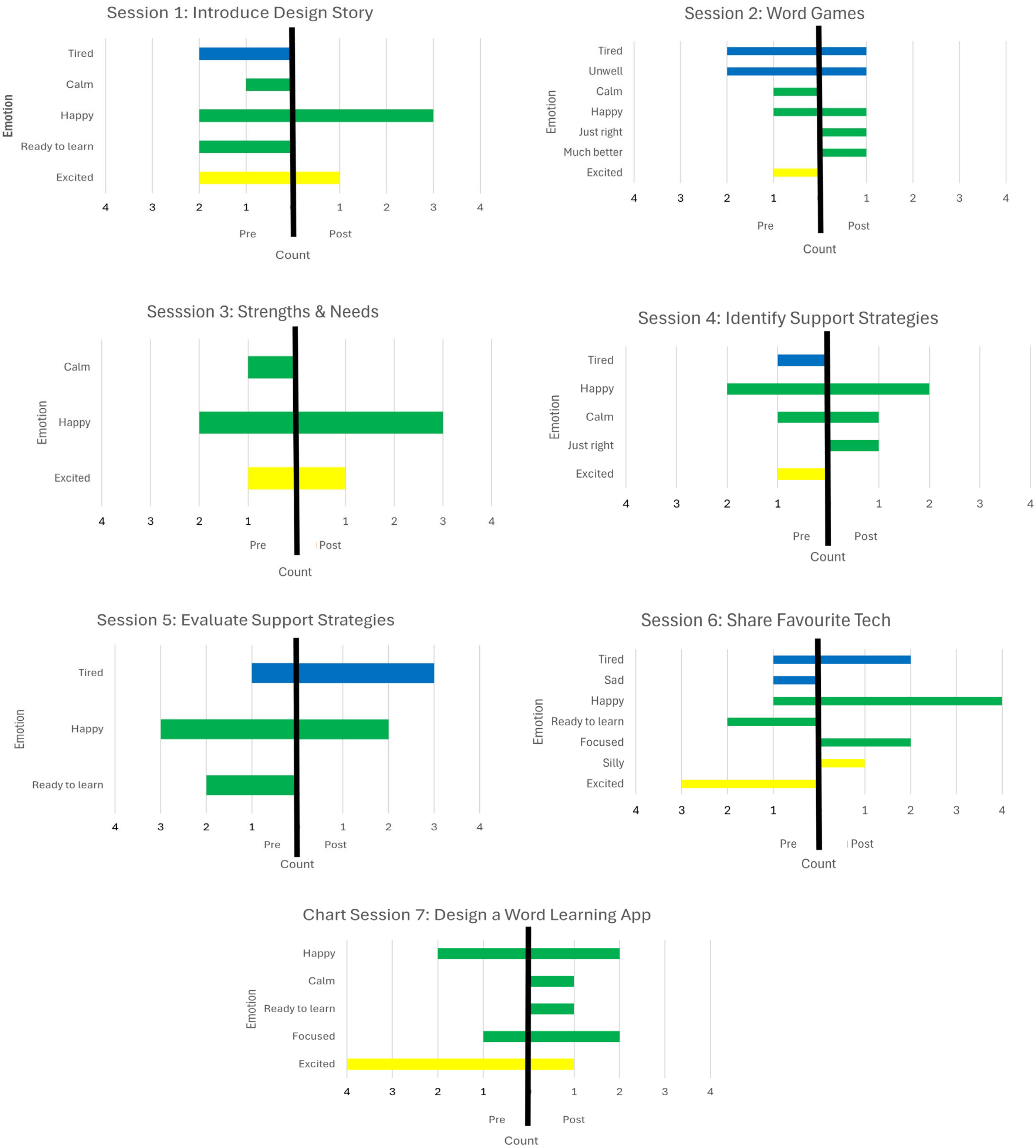

Figure 3 Children's emotional responses before and after each design session
Based on The Zones of Regulation™ approach [48]: blue emotions represent low alertness and down feeling; green emotions represent calm state of alertness ideal for engagement and learning; yellow emotions represent heightened state.

**Theme 2: Stories incorporating familiar and engaging characters can make design sessions meaningful and interesting**

The first participatory design session (Introduce Design Story) focused on the Cooperative Inquiry technique of introducing the project to the children as a story [21,24]. The children were informed that we would be designing a word-learning app to help superheroes with using the right words. Superhero props e.g. comic books, masks, pictures, were provided as visual prompts which were brought back in subsequent sessions. The children appeared interested in the story idea and enjoyed exploring the props (Fig 4). Written feedback indicated that the children responded well because familiar characters that they enjoyed were used to set the scene for the design sessions:

> '*I like the superheroes because we were seeing real superheroes.'* (Child 2)

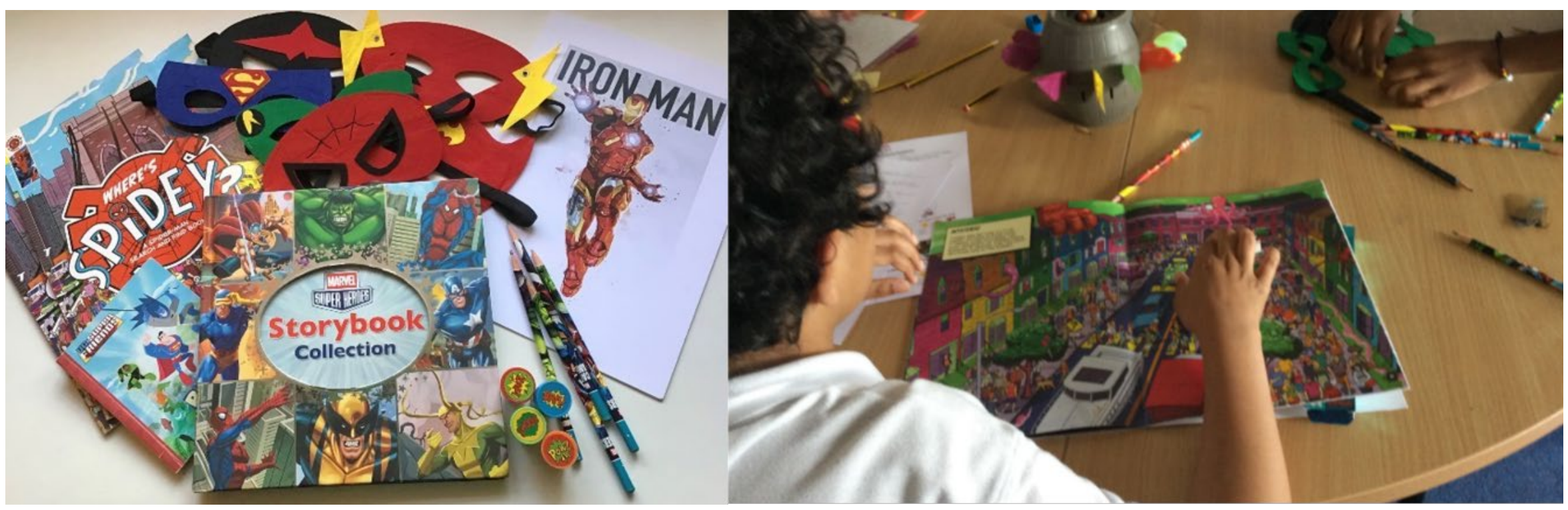

Figure 6.4 Children engaging with superhero props for storytelling in design sessions

Emotions selected by the children following session one also indicated a positive experience, with three children selecting '*happy'* (Child 1, 2, 3) and one '*excited'*. This is despite half of the children, stating they were '*tired'* pre-session (Child 1, 4).

The same books and pictures were used in subsequent design sessions which led to ongoing engagement. For example, in session two (Word Games) superhero word games helped explore children's responses to vocabulary activities. The children engaged well with the design activity which involved generating their favourite superhero words. They produced a large number and range of words, some relatively complex considering the children's underlying language difficulties, e.g. '*holograph', 'claw'* and '*shield'* (Fig 5).

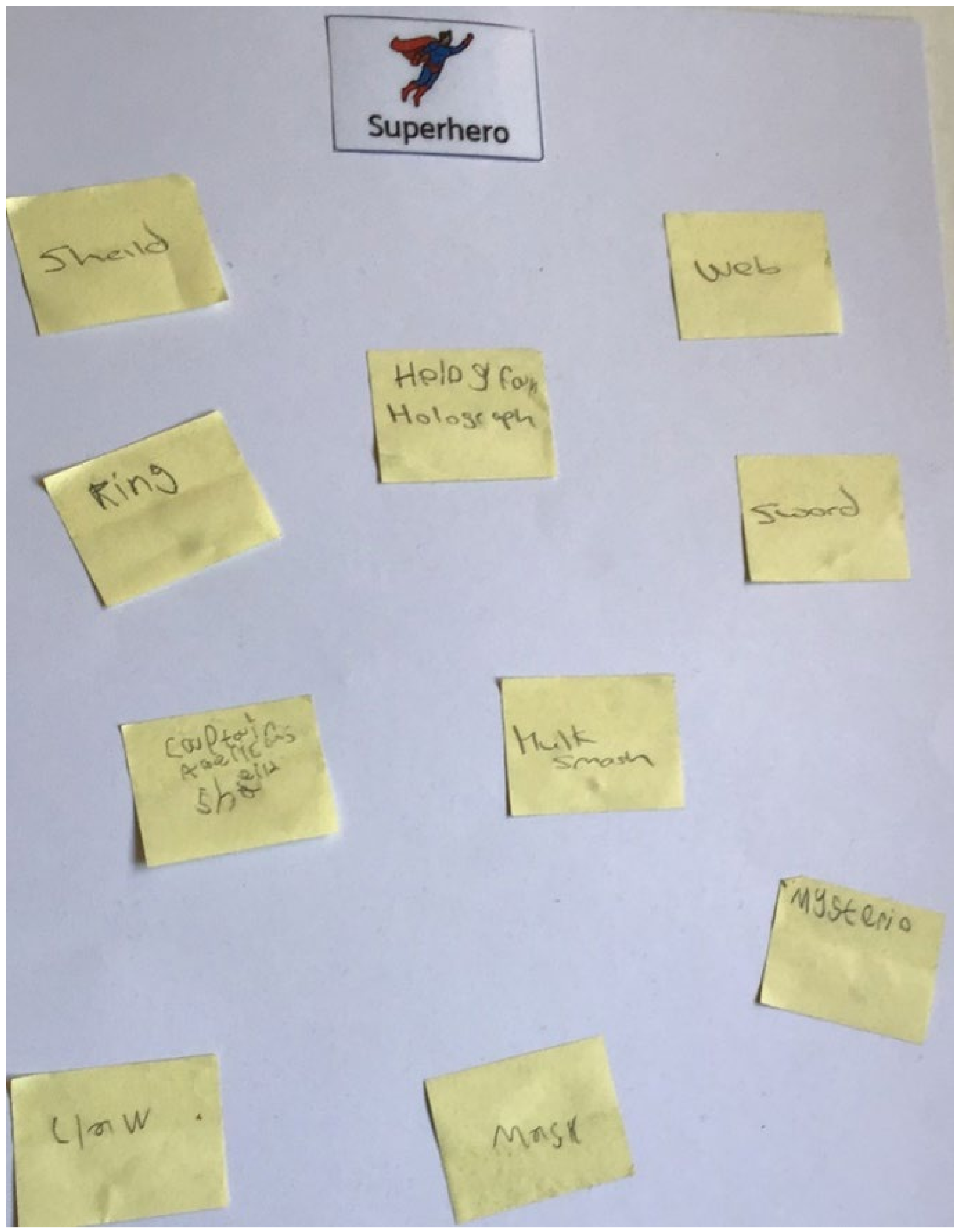

Figure 5 Design activity with children generating high interest words (superhero theme).

**Theme 3: Introducing preferred videogames can help children relate to design activities with confidence but may also lead to heightened emotions**

Session six (Favourite Tech) adopted the Cooperative Enquiry technique of 'Technology Immersion' to understand children's tech preferences [21]. The children engaged with two familiar videogames Minecraft and Dream League Soccer 2024 - DLS24. Pre-session, children's emotions were mostly '*excited*' (Child 1, 3, 4) followed by '*ready to learn*' (Child 1, 2) indicating a positive anticipation of what was to come. The children were observed to know the games well and appeared more confident during this session compared to the previous sessions where videogames were not used. The session also allowed the children to show strengths, interests and aspirations which they may not have otherwise shared.

> *'My dream job is a Minecraft You Tuber.'* (Child 3)

All the children stated they felt '*happy*' after the activity. In addition, '*tired*' [Child 1, 2] and '*focused*' (Child 1,4) were selected in equal numbers. Of note, one child (Child 2) selected '*silly*' as one of their emotions after the session and in line with the Diversity for Design approach [22] was supported to regulate by their TA using a self-selected squeeze toy before returning to class.

The confidence observed in the 'Technology Immersion' session continued in session seven (Design a Word Learning App) where the children were given arts and crafts to generate design suggestions based on a Cooperative Enquiry approach known as 'Bags of Stuff' [21]. In continuation with the fictional narrative theme, the children were asked to design a game for their character of choice. All the children selected '*excited*' at the start of the session, two stated they were also '*happy*' (Child 2, 4) and one also selected '*focused*' (Child 1).

Children were observed engaging well throughout the activity and enjoyed taking on the role of an app designer:

> '[I liked] *when we were designing a game and who we are going to help.'* (Child 1)

All the children selecting positive emotions following the session: '*happy*' (Child 1, 2), '*focused*' (Child 1, 4), '*ready to learn*' (Child 1) and '*calm*' (Child 3).

**Themes 4: For a positive experience, the processing demands of design activities need to align with children's capability and capacity**

There were occasions when children were unable to participate due to the demands of the design activity. For example, session four (Identify Support Strategies) aimed to enable children to inform the design by selecting preferred language support strategies (Informant Design, [20]). However, barriers to involvement were noted such as not all children feeling comfortable speaking during group activities, and some children still having more to say but not having the time. In order to support, post-it notes were placed next to all the children to use instead of/in addition to verbal participation. This strategy was successfully adopted by all the children and emotions post-session were '*happy'* for three of the children (Child 1, 2, 3) and '*excited'* for one (Child 4). The use of post-its was therefore introduced as standard in subsequent sessions.

Session five (Evaluate Support Strategies) adopted the Informant Design [20] approach to understand why the children preferred particular language-support strategies. The activity involved the children evaluating the word-learning strategies they currently use, something they struggled to do. More specifically, the children hesitated in using the feedback sentence prompts (Fig 2) to describe what they liked/disliked/would change about a strategy. They either opted out or stated, *'I don't know'.* The children's emotions post-session also indicated that the session was demanding, with three of the children selecting '*tired'* (Child 1, 2, 3) even though only one child had selected '*tired'* pre-session (Child 3). The findings were reflected on by the primary author (RA) with the school support staff and it was agreed that in subsequent sessions children would not be asked to evaluate language strategies and instead adult observations would be used to understand why children preferred some strategies to others.

**Summary of children's participatory design experiences**

As a group, the children demonstrated patterns of responses to certain design activities. In particular, a positive reaction to familiar story characters and video games and to providing design ideas. Situations with restricted options for feedback (e.g. spoken only or in a group) were more challenging but when more choice was given engagement improved. The language and cognitive demand of activities affected responses, with higher demand leading to reduced interactions.  Within the responses nuances emerged such as highly anticipated and exciting activities may lead to tiredness.

## 3.2 What are the system requirements when designing a word learning tool for neurodiverse children?

In addition to informing children's design experiences, the data generated from the sessions were analysed thematically to provide insight into requirements for creating a word-learning intervention that would engage and sustain children's motivation.

**Theme 1: Children are responsive to multimedia opportunities for gaining and sharing information but struggle with abstract content**

A theme apparent across the design activities was that children would engage better with a system that could provide a range of multimodal learning and response opportunities.

This was evident from the first session with children using a range of multimodal communication methods to provide responses during design activities including spoken, written, drawings, photos and videos. Some children did indicate a preferred communication method (e.g., Child 3 spoken), however, most combined multimodalities such as creating videos with a spoken narrative (Child 4).  Preferences varied depending on factors such as the nature of the design activity, the resulting interactions, and the child's emotional state (e.g. Child 1 was more likely to provide written feedback when tired).

Children's reactions to multimodal stimuli were also informative and highlighted a preference for relatable and interpretable content. This was evident in responses from session six (Favourite Tech), where children compared videogames DLS24 and Minecraft. DLS24 uses realistic, recognisable sound effects and music that map clearly onto actions and meanings. In contrast, Minecraft incorporates a deliberately abstract and ambient sound design, including non-literal environmental sounds and distorted effects. These occur sporadically and unpredictably and are not consistently linked to gameplay events, time, or location, instead serving to create atmosphere rather than convey clear meaning [46]. Some children appeared to struggle with this abstraction, describing the sounds as confusing or unsettling rather than supportive of understanding:

*'I don't like the creepy sounds. They don't make sense.'* (Child 2)

In DLS24, the authenticity of the graphics in DLS24 was seen as a further positive:

'*Loved it. The players look so real. So sick.'* (Child 1)

**Theme 2: Customising learning to children's interests encourages a positive attitude**

A particularly informative interest-led learning theme emerged in session three (Strengths & Needs). Responses for children's attitude towards their strengths indicated that the children related strengths to things they are interested in:

*'My strengths are my favourite things; they are the things I am good at.*' (Child 1)

'*Your strengths are the things that you like.'* (Child 2)

Children's responses to the vocabulary activity in session two (Word Games), also indicated that the more the children were interested in the learning the better the engagement. In the activity, the children selected favourite superhero characters and generated as many favourite superhero 'words' as they could think of*.* The children were observed engaging well and enjoying the superhero-themed vocabulary activity producing a large number of words (Fig 5).

Children were also more likely to show interest in learning if they had the opportunity for customisation. This was seen in session seven (Favourite Tech) which generated design requirements by understanding the children's technology preferences. The two videogames trialled by the children allow a high degree of individualisation, Minecraft encourages individual creativity when crafting worlds and DLS24 enables users to create their dream team by selecting players and personalising kits, logos, and stadiums. This tailoring to individual preferences appealed to the children:

*'I like you can customise it* [DLS24].' (Child 1)

*'Look what I made! It's mine!* [Minecraft].' (Child 3)

**Theme 3: Repeated, child-led learning opportunities promote skills**

In session four (Identify Support Strategies), factors important for learning were explored. Of note, children identified the chance for multiple practice as important. More specifically, children described repeated learning opportunities as a way to develop a skill, a feature that can be embedded across the system:

*'The more you do, the better you get.'* (Child 3)

A corresponding theme emerged in session seven (Design a Word Learning App). Children reinforced the benefit of repeated practice but highlighted the importance of having good instructions.

*'In the first screen we need a pop-out with the rules, so we know every time what to do.'*

**Theme 4: Learning through technology must be embedded in the real world to complement existing practice and align with children's individual circumstances**

Children's preferred language support strategies as explored in session four (Identify Support Strategies), indicated that the people supporting them, and the mood of the children were important considerations. When the children were asked for strategies to support word learning, multiple suggestions were generated including directly learning about words as would be expected (e.g. using dictionaries, language worksheets, word webs). However, suggestions extended beyond target words, with the children also highlighting the importance of adult support:

*'I would ask a teacher for help.'* (Child 3)

*'You can ask Google, but you can't ask Google all the time. You need to ask people. People know things.'* (Child 2)

Children also referred to factors to support word learning that related to their overall state and the importance of timing:

*'Wait till they're ready to listen.'* (Child 4)

As well as general wellbeing strategies:

'*Taking deep breathes helps.*' (Child 1)

Another finding related to children connecting with their surrounding environment during word learning emerged in session seven (Design a Word Learning App). One child (Child 4) used roleplay and props to demonstrate how augmented reality (AR) could create a bridge between the digital intervention and the real world. He chose the word 'football' to target and started by drawing an iPad then used it to 'record' him kicking a football to another child, whilst giving a description of how the ball was being used (Fig 6a). The child made reference to Pokémon Go [Niantic, Inc, www.pokemongolive.com], an AR mobile game that combines the real world with the virtual Pokémon universe [Fig 6b)

*'There should be cartoons and then real things, like in Pokeman Go.'* (Child 4)

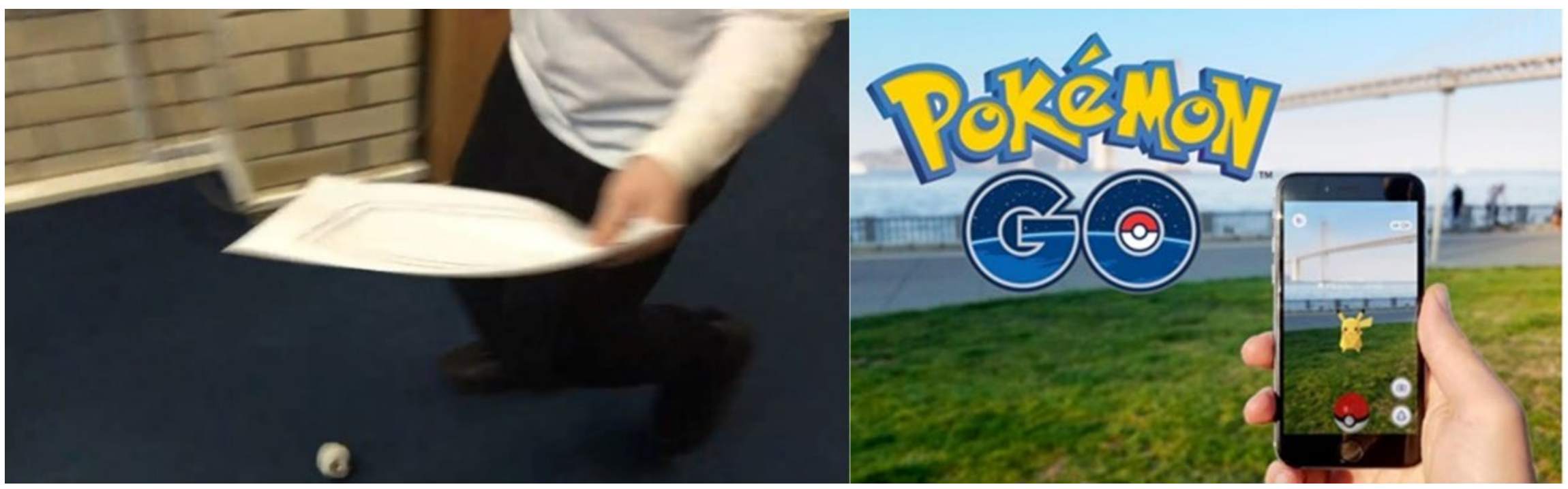

Figure 6 Augmented reality design ideas
Fig 6a Left: Child 4 engaged in participatory design role play where he drew an iPad to 'capture' a football [target word] from the real world into a word-learning app. Fig 6b Right: The augmented reality game Pokémon Go (Niantic, Inc, www.pokemongolive.com) inspired Child 4's design ideas.

**Theme 5: Playfulness and error-less learning are key motivators**

A notable theme from session six (Favourite Tech), was an enjoyment of open-ended play such as in Minecraft because the focus on exploration and creativity meant it was perceived as a 'no-fail' experience:

*'No such thing as game over!'* (Child 2)

Children's existing positive play experiences with technology also informed their suggestions for designing a word-learning app in the fourth design phase. The children referred to the need for tracking progress within the app. This theme of progressing through the app was inspired by the children's gameplay experience:

*'There should be levels, like your average video game.' (Child 3).*

One child described a visual bar to show how much work had been completed:

*'When you do more learning, you fill more of the bar, and this shows how well you're doing.'* (Child 2).

Another child drew a points chart where children could earn points as they progressed through the app (Fig 7), the description they wrote reinforced the 'no-fail' theme as rewards were linked to completing more of the app rather than a change in difficulty:

*'Game hardness is the same.'* (Child 1)

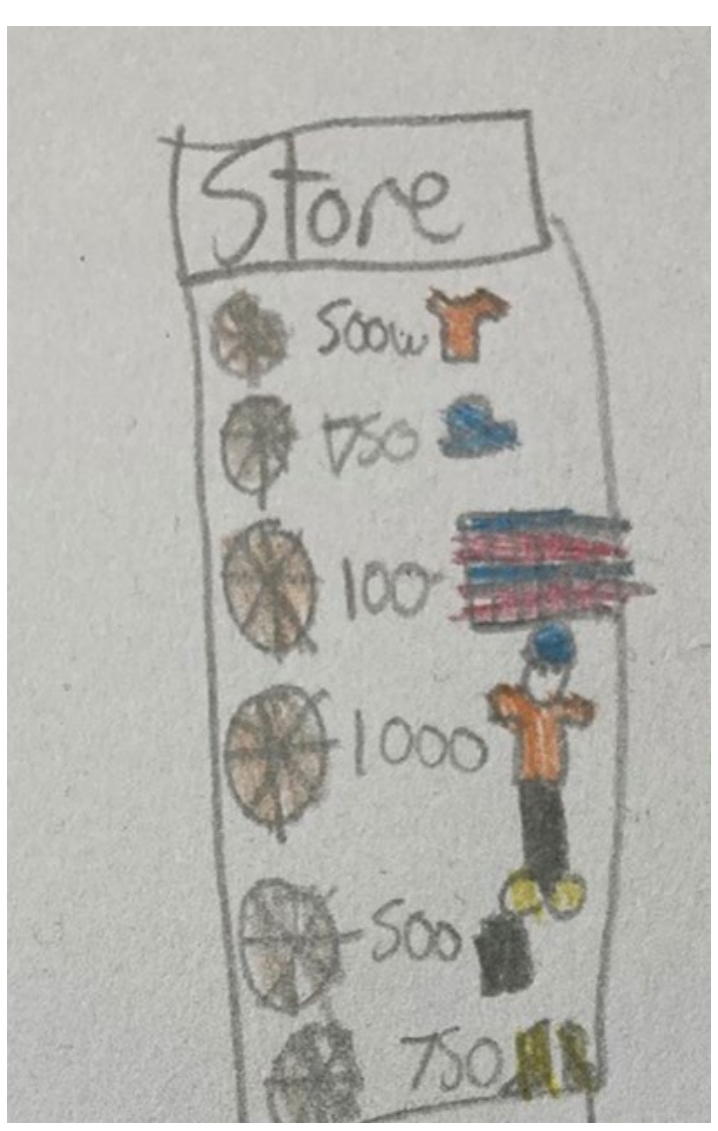

Figure 7 Drawing of a reward system for the word-learning app

**Summary of Findings and Implications for App Development**

The children provided clear indications of design requirements for the end-solution to be engaging. This included a choice of multimodalities when accessing and providing information. Repeated learning opportunities, indications of progression, and customisation features were found to be motivators. Alignment of real and virtual learning components was important, such as adults supporting the children in-person during word learning and the system combining a cartoon-style interface with uploaded media from the child's environment.

As summarised in Table 5, the children's design requirements were translated into intervention design features and a prioritisation matrix was applied. An adaptation of the Eisenhower Method [47] was used, whereby each design feature was rated according to its value to the user (children with DLD) and technical feasibility. The next step will be to coordinate the word-learning app design requirements and proposed features for child engagement and motivation with corresponding therapeutic content requirements which have been guided by word-learning theory, research, and clinical practice [30]. Integrating these engagement and therapeutic requirements will establish readiness to progress to app development.

Table 5: Prioritisation matrix to inform intervention design choices

| Child Requirement | Intervention Design Features (IDF) | Value | Ease of Feasibility |
|---|---|---|---|
| Theme 1: Children are responsive to multimedia opportunities for gaining and sharing information but struggle with abstract content.<br>- Children use and respond to multiple communication methods depending on personal preference and activity type.<br>- Better response to relatable/concrete multimodal stimuli than obscure/abstract media (preference for sounds that matches the scene). | **Multimedia support** (text, audio, image, video) that is user-centred and coordinated to complement learning | High | High |
| Theme 2: Customising learning to children's interests encourages a positive attitude.<br>- Children relate strengths to things they are interested in.<br>- Greater interest in customised activities. | **Customised** learning, using children's own photos, sounds, or videos for target words* | High | High |
| Theme 3: Repeated, child-led learning opportunities promote skills. | **Child-friendly interface with consistent prompts** for progression from adult-facilitated to self-directed use | High | High |
| | **Child-directed tutorials,** in addition to tutorials for supporting adults, that can be accessed repeatedly. | High | High |
| Theme 4: Learning through technology must be embedded in the real world to complement existing practice and align with children's individual circumstances<br>- Children value adult support during word learning.<br>- Children included word-learning support strategies relating to their general state (readiness for learning)<br>- Children find linking the real and digital world engaging (augmented reality, mixing real media and cartoon style) | **Prompts** where adult can interact with child to provide social support and promote blended learning | High | High |
| | **Introduce Augmented Reality features** | Medium | Low |
| | **Cartoon-style design features** for accessibility and engagement. | High | High |
| Theme 5: Playfulness and error-less learning are key motivators.<br>- Open-ended exploration is appealing<br>- Indicators to track progress are appealing | **Child-directed progress indicators** reflecting level of engagement. | High | High. |

## 4 Discussion

This paper reports on the experiences and design requirements of children with DLD when a bespoke participatory design approach was used to inform a word-learning intervention.

### 4.1 Design experience

The implications of children with DLD participating in an accessible, collaborative, and learner-centred design approach are discussed below.

**Identifying and understanding how different factors may influence children's design experiences**

The Diversity for Design framework [22] provided a wealth of additional relevant information relating to the child beyond the design activity itself. This included pre-consideration of the children's strengths, needs and, preferences as well as their self-reported emotional state pre-session, whilst also having staff familiar to the children to refer to during the sessions. All of this provided extensive information of factors important to the children from which more isolated findings from specific design activities could be considered.

Influencing factors beyond the design sessions are particularly important for children with DLD as presentation often varies depending on the extent to which their external environment serves either to support or to add demand [48]. As discussed by Frauenberger et al. [49] in their paper on rigour and accountability in participatory design research, clear reporting of contextual dependencies is important to enable robust interpretation and generalisation of findings.

**Explicit value of the design experience for the children**

The success of the fictional narrative technique from the Cooperative Inquiry approach [21] in providing collaborative coherence was due to the familiarity of the characters. As popular superheroes were used, the children were engaged and motivated to help these characters by designing a word-learning app for them.

This raises important points about the purpose and value of the design sessions from the children's perspective and how they may differ from those of the researchers and designers. As explored by Spiel et al. [49] realisation of anticipated benefits of participatory design outputs for the children involved in the research projects tends to be rare and so explicit situational benefits should be incorporated. In this study, one explicit benefit for the children which was realised within the design sessions was the opportunity to enjoy superhero stories.

**Empowerment through participatory design**

Another direct benefit from the design sessions for the children, facilitated by the Cooperative Enquiry approach of children being collaborative design partners [21], was a sense of empowerment. This has been defined in the participatory design literature as the extent to which children care about the activities and their perception about the relevance and importance of their contribution [23].

The children enjoyed playing video games during design sessions, they also shared certain knowledge and skills which they are not generally able to demonstrate in school. They then applied their videogaming expertise as app designers in the final session, and as a group identified as feeling calmer, more focused and with greater readiness to learn compared to before the session. This indicates that being empowered to express strengths during participatory design may have resulted in the children returning to class with a more positive and facilitative attitude to learning. The finding has parallels to that of Malinverni et al. [23] who, when designing a virtual reality social skills tool with neurodiverse children, reported that the children were keen to share their ideas with their parents after the session and to develop videogames at home.

**The risk of overburdening**

Working with children to solve design problems inevitably places a burden on them. For an empowering participatory design experience, the demands of this burden must be within the child's capacity and capability. Adopting the Diversity for Design approach [22] and gaining regular feedback from the children highlighted times when this balance was not achieved and methodological changes needed.

It is worth noting, that even experiences which the children reported as enjoyable could have been burdensome. For example, whilst most children selected '*happy'* after the session playing videogames, half also selected '*tired'* and one selected '*silly'*. Given that design research in a child's routine environment is advocated [20-22], there needs to be consideration of the support required not just in the design sessions but also during the return to school routine.

**Recommendations for participatory design with children who have DLD**

The richness of the collected data enabled children's experiences to be continually reviewed and adjustments made to design activities prior to subsequent sessions. An increasingly bespoke design approach evolved, aligned to children's strengths and preferences, something that is highly advocated by the wider literature as it represents best practice [28-29]. Considerations from this study have helped create recommendations for involving children with DLD in participatory design as summarised in Table 6.

Table 6 Recommendations for involving children with DLD in participatory design

| | |
|---|---|
| 1. | Create bespoke participatory design experiences by using each child's presenting profile to individualise established design templates and frameworks. |
| 2. | Work with the children and their support professionals to understand important factors outside of the design session that may influence children's design experience. |
| 3. | Ensure explicit situational benefits of participatory design for the children such as access to favourite stories and games. |
| 4. | Continually review and respond to children's experiences to empower without overburdening. |
| 5. | Participatory design is an ideal method to identify engaging and motivating design features, however, don't expect all design questions to be answered using this approach. Instead, plan according to each child's interest and capacity, and seek other research methods for outstanding queries. |

## 4.2 Design requirements for a word-learning app

A range of intervention requirements were generated by the children, giving insight into the design features needed to make the digital word-learning engaging and motivating for children with DLD as discussed below.

**Use of multimedia to provide choice of multimodal learning experiences**

Children responded positively to a choice of multimedia for accessing and responding to information indicating that this is an important design feature. Wider literature suggests that consideration must be given not only to choice of multimedia but also how these can be coordinated to enhance learning.

This is effectively captured in the work by Mayer [51] who proposes a Cognitive Theory of Multimedia Learning and provides corresponding instructional design principles. The design principles optimise cognitive processing during multimedia learning by (i) minimising extraneous processing demands such as distracting background noise, (ii) facilitating intrinsic processing through accessible and manageable tasks, and (iii) optimising novel processing using cues to align acquired and existing knowledge.

The principles have been adopted by Shaban and Pearson in their Learning Design Framework [52] which aligns Mayer's work with user-centred principles specifically for children with learning disabilities. For example, reducing extraneous processing demands by highlighting information in the app that is important for the child. The framework by Shaban and Pearson and its application to intervention design have been validated and evaluated empirically [53-54], making it an ideal guide for coordinating multimedia within the word-learning intervention.

**User-generated content**

Personalisation of systems emerged as important during the design sessions. This could be achieved in the intervention by children creating their own word-learning content as has previously been explored in an intervention app for children with ASD [15, 55]. That app allowed children to insert their own photos into a dictionary format and add audio recordings alongside the written word to build a personalised catalogue of words over time. The design was influenced by discussions with parents, teachers and SLTs as well as observations of children with ASD being engaged and motivated to communicate when sharing photos they had taken on their iPads.

When tested by teachers and SLTs, reported benefits included additional learning opportunities from children creating their own multimedia rather than depending on pre-loaded, pre-selected images. Children's agency and independence were reported to build over time '*That's the first session that Child F has just gone off on his merry little way. Prior to that it was very adult-driven "What are you going to take a photo of next?". So, they're starting to come up with their own ways of using it I think*' (Teacher).

A limitation of the app was the lack of a video function which the investigators concluded would have facilitated more dynamic learning opportunities.

**Repeated, child-led learning opportunities**

Children's emphasis on repetition, together with their request for clear and consistently available instructions, highlights the importance of embedding repeated word learning opportunities supported by accessible guidance. This aligns with wider evidence that children with DLD typically benefit from increased opportunities to retrieve and use newly learned words, rather than relying on single exposures. Experimental word learning studies show that spaced retrieval practice can improve word learning outcomes for children with DLD, with repeated spaced retrieval supporting longer term recall [56, 57]. While the DLD specific digital vocabulary evidence base remains comparatively limited, the available research refers to positive word learning outcomes being linked to digital formats enabling repeated access to structured learning opportunities [9, 10]. Taken together, the findings and existing evidence indicate that the word-learning intervention should support repeated practice through a child-friendly interface that provides clear, consistently available prompts and guidance that can be accessed flexibly.

**Technology embedded in the real world**

The design sessions helped to understand environmental influences on how the word-learning intervention might be used. For example, in addition to child-directed instructions within the app, the continued input from supporting adults to complement the digital resource was considered important to the children, as per a blended-learning model [11].

Suggestions were also made on the use of augmented reality [AR] to connect the real and digital word. This has been previously researched by Shemy et al. [58] who gained design ideas for how AR could provide additional contextual information to enhance a word-learning intervention by conducting focus groups with experts and interactive workshops with ASD children and their families. SLTs, psychologists, and special education teachers provided understanding of how AR could align with word-learning theory and practice. Child observations provided additional information by indicating greater engagement when cartoon animation augmented realistic elements from the environment, which complements design ideas from this study.

**Playful Learning**

This study found that playful learning opportunities were likely to increase intervention engagement and motivation. For example, children showed a particular preference for the Minecraft videogame because it enabled error-less exploratory play**.** Given that language acquisition requires rich exploration of words and word features [59], techniques from exploratory gameplay could help inform the word-learning intervention.

Another gameplay feature that children expressed a preference for was the use of levels to demonstrate progression. It was important to the children that this was based on how many activities were completed rather than rather than how well they performed. This aligns with findings from Huang et al. [60] who analysed gameplayer behaviour and identified that indications of influencing a system was important for high engagement, while the ongoing sense of being challenged discouraged continuous play.

**Recommendations when designing a word-learning intervention for children with DLD**

Gaining the views of children with DLD through participatory design has highlighted intervention requirements that are of importance and value. Aligning these with existing intervention-design theory and research, has helped create a series of recommendations to ensure that the word-learning intervention is sufficiently engaging and motivating to be accepted by children with DLD as summarised in Table 7.

Table 7 Recommendations for designing a digital word-learning intervention for DLD children

| | |
|---|---|
| 1. | Coordinate multimedia options including text, audio, images, and video to provide a range of multimodal learning opportunities. |
| 2. | Encourage meaningful, exploratory learning by enabling children to create media from their own interests, routines, and environments. |
| 3. | Introduce repeated, child-led learning opportunities to promote skill retention |
| 4. | Facilitate blended learning through a child-friendly interface that directs the child as the learner whilst also being relevant to the adult supporting the learning |
| 5. | Demonstrate progress made using visual indications of level of engagement rather than performance. |

## 4.3 Reciprocity between design experience and design ideas

Design experience and design ideas, whilst being distinct entities, are interdependent. As noted in this paper, the better the design experience for the child, the more likely they are to engage in generating design ideas.

Furthermore, the learning that came from understanding the children's experience of the design sessions provided valuable insight into the requirements of the system being designed. For example, understanding factors outside of the design session that affected the children's design experience indicated the external factors that may influence how a user would engage with the designed system and possible mitigations. Similarly, the response of the children to multimodal opportunities for receiving and providing information during the design sessions, helped understand the role of multimedia in the intervention design.

# 5 Limitations

Children shared their experience of participatory design using emotion symbols. This is an established strategy to support children with DLD that the children in this study were familiar with [61]. However, children may have been influenced by the selection of emotions presented, and an attempt to mitigate this was made by encouraging them to consider other emotions, which they did at times.

Another limitation was that, whilst professionals played a significant role in the participatory design process, interactions with parents was more limited. Parents did contribute to the children's 'About Me' profiles which helped individualise the design sessions. Children also routinely took home artefacts from the sessions such as drawings to share with their family, and general updates were provided through emails and group meetings. However, given the unique perspective that parents have on the impact of DLD on their children [62], the intervention design will benefit from more established methods of parental involvement, such as interviews and focus groups. This will be an area of future research.

## 6 Future Work

The child engagement and motivation requirements for a digital word-learning tool intervention as identified in this study will be considered alongside theory, evidence and current practice around supporting children with DLD [see 30]. This will collectively inform the development of the system, which will then be evaluated for usability.

## 7 Conclusions

Given the benefits for the children involved in the design process, and the benefits for the system being designed, it is advocated that investment in understanding and responding to children's design experience should be considered best practice in the field of participatory design.

Whilst the focus of the work has been children with DLD, the widely applicable nature of findings and recommendations suggests that they are likely to be of interest to researchers, clinicians and designers supporting a range of neurodiverse children.

## 8 Declarations

### Conflicts of interest

The authors declare no conflicts of interest.

### Data availability

The underlying study data are not publicly available because the small sample and detailed qualitative material could make participants identifiable. De-identified excerpts supporting the findings are included in the article.

### Author contributions

**Dr Rafiah Ansari:** Conceptualization; Methodology; Software; Formal analysis; Investigation; Resources; Data curation; Writing – original draft; Writing – review and editing; Visualization; Project administration; Funding acquisition.

**Dr Rina R. Wehbe:** Methodology; Validation; Resources; Writing – review and editing; Visualization.

**Dr Lizbeth Olivia Escobedo:** Methodology; Validation; Resources; Writing – review and editing; Visualization.

## Appendix A: 'About Me' profile form

Completed by child participant with support from parents, teachers and school speech and language therapist

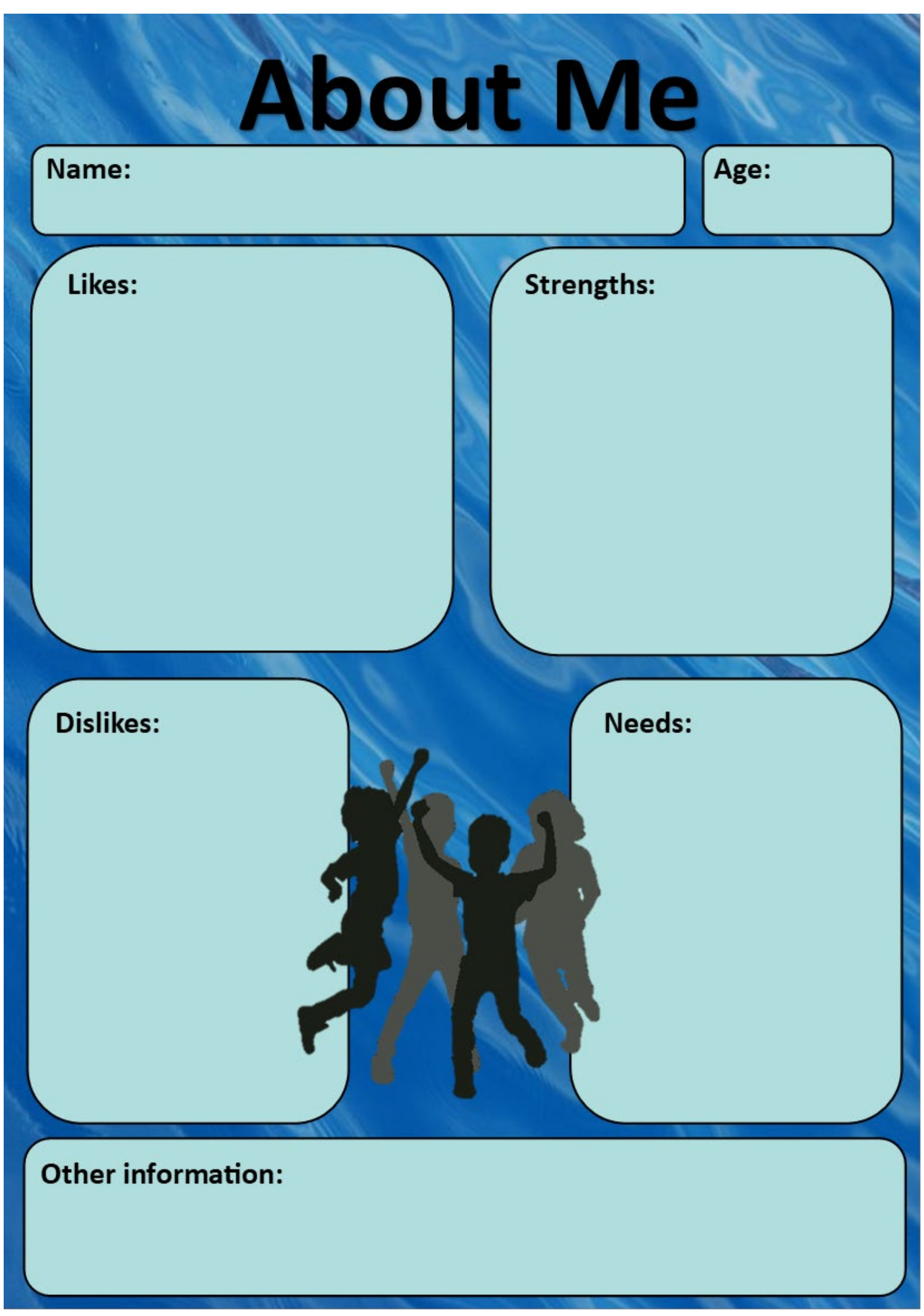

## Appendix B Further details of qualitative analysis of children's responses (codebook, themes and subthemes)

Thematic analysis of study data

| RQ1 | Themes | Subthemes | Corresponding content from the data (text used in final manuscript highlighted in bold) |
|---|---|---|---|
| What are the experiences of children with DLD when participating in an accessible, collaborative, and learner-centred design approach? | 1. Exploring children's feelings at the start of design sessions indicates factors that may influence children's design experience and readiness for design activities. | Tiredness as design sessions were after lunch gives insight into factors influencing children's capacity during the school day. | Fieldnotes (Session 1 – Design Story): Following the session the Principal Researcher (PR) explored with SENCo and SLT that half the children chose tired at the pre-session emotion check-in (SLT had been present in session with the children's teaching assistant - TA). SENCo advised that no other session slots were available. Both SENCo and SLT felt the children's mood provided a realistic picture of changes in capacity that the intervention being designed would need to account for. Also of note, none of the children selected tired post-session.<br><br>Emotion scale (Session 1 - Design Story):<br>Child 1: Pre-session - Tired, Excited, ***We had lunch and so I'm sleepy but I want to join.*** *I'm excited in the Rainbow room*. Post-session - Happy<br>Child 4: Pre-session - *Same as* (Child 1). Post-session - Excited |
| | | The sessions gave opportunity for children to explore their emotions with support staff and were then given the agency to decide whether to participate. | Fieldnotes (Session 2- Word Games): Partnership model (Cooperative Enquiry, Guha et al., 2013) applied when the children selected a 'blue' emotion on the emotion scale pre-session (Zones of regulation, Kuypers, 2011, blue for low alertness and down feeling - sad, tired, unwell, bored). We only continued if they and their TA/teacher were in agreement. |
| | | | Emotion scale (Session 2 - Word Games):<br>Child 2: Pre-session - Calm, Unwell. ***I have a cough.*** Post-session -<br>Child 4: Pre-session - Unwell. ***I have a cold.*** Post-session -<br><br>TA Comment*: I've checked with the boys and their teacher, both want to do the session and the teacher's happy* (Session 2 – Word Games). |
| | | | Emotion scale (Session 6 - Favourite Tech):<br>Child 2: Pre-session – Sad. Post-session – Happy, silly<br><br>TA Comment*: I've chatted with (Child 2), he's ok. He had a tricky time in the playground. His teacher helped with the incident, we both think he'll be ok and I'll keep an eye on him* (Session 6 - Favourite Tech). |

<table>
<tr>
<td rowspan="2"></td>
<td rowspan="2">2. Stories incorporating familiar and engaging characters can make design sessions meaningful and interesting</td>
<td rowspan="2">Good engagement with the ideas of designing a word-learning app to help familiar superheroes with using the right words.</td>
<td>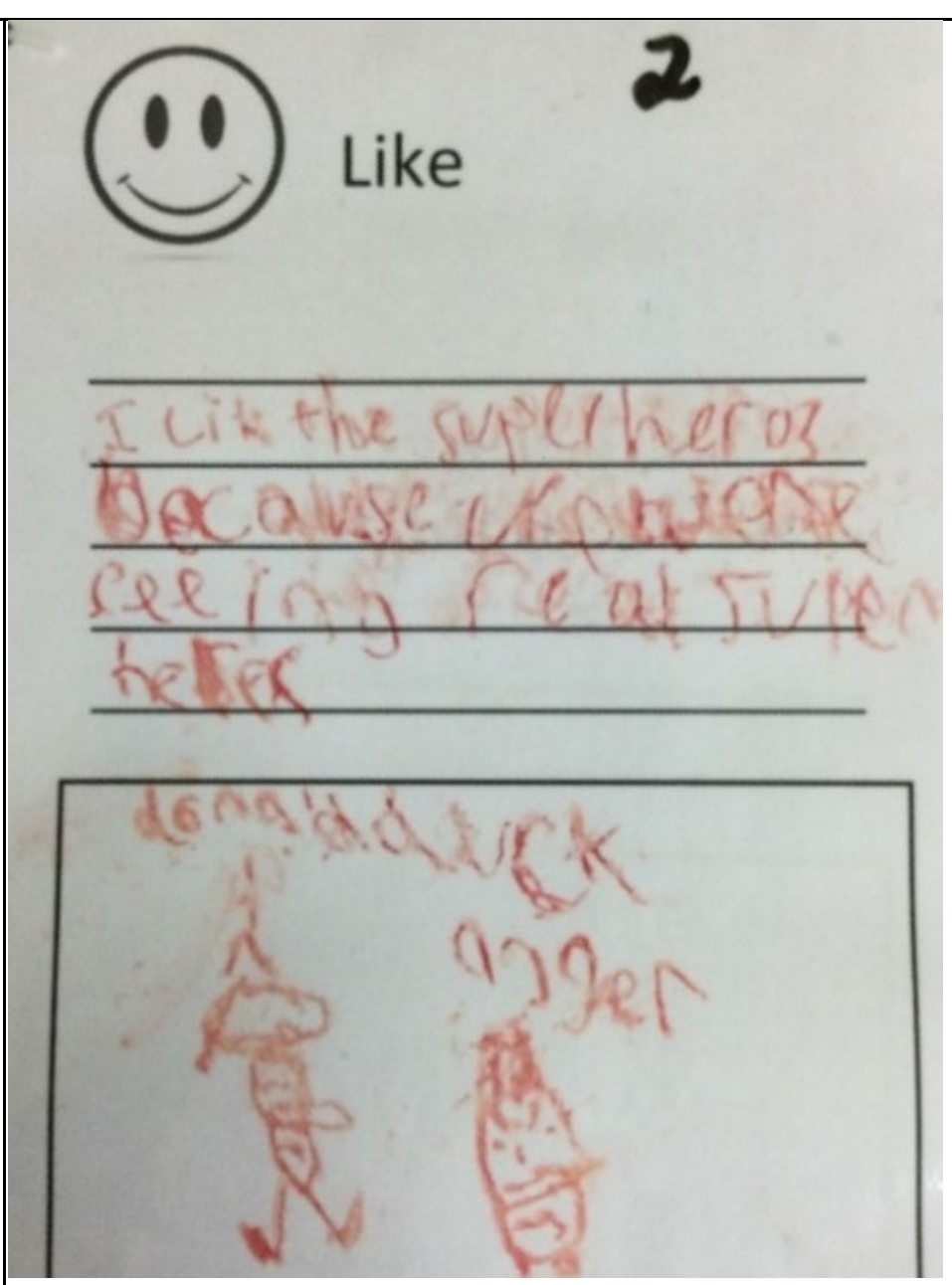
</td>
<td>Written/drawn feedback<br>(Child 2, Session 1 - Design Story)<br><br>***I like the superheroes because we were seeing real superheroes***<br><br>*[Text illegible]*</td>
</tr>
<tr>
<td colspan="2">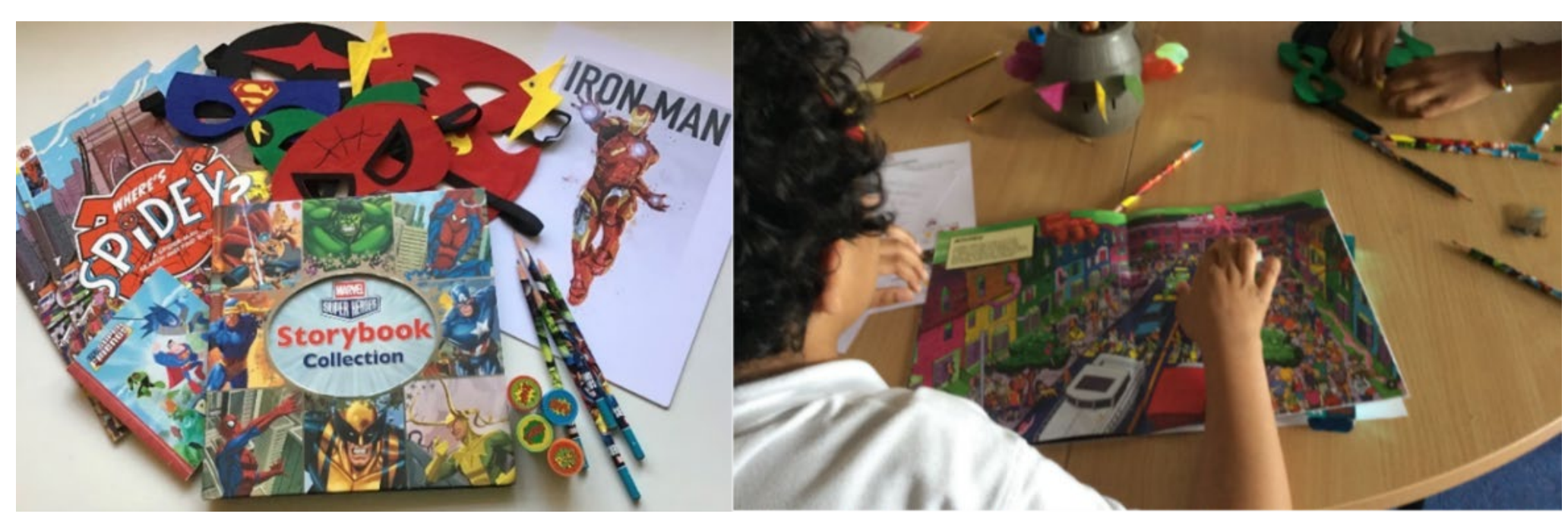

Photos of superhero props used for storytelling in design sessions and children engaging with superhero props (Session 1 - Design Story)<br><br>School SLT comment: *They did great! I'm surprised by some of the words and how long they kept going for. They loved the superhero idea. Really pleased* (Session 2 - Word Games)</td>
</tr>
</table>

| | | | | |
|---|---|---|---|---|
| | | | 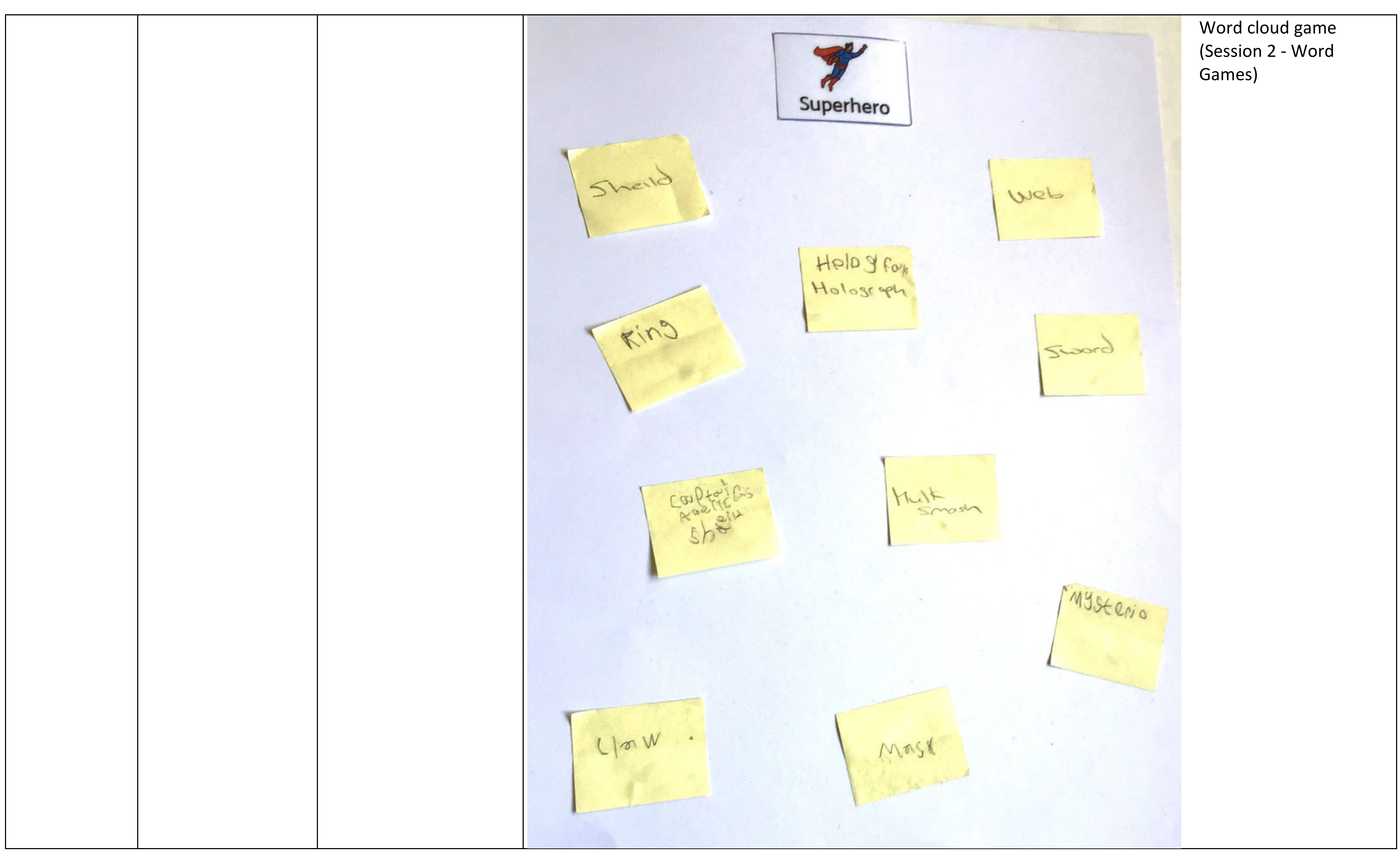 | Word cloud game (Session 2 - Word Games) |

| | | | |
|---|---|---|---|
| | 3. Introducing preferred videogames can help children relate to design activities with confidence but may also lead to heightened emotions | Children showed strengths, interests and aspirations they may not have otherwise shared. | Fieldnotes (Session 6 - Favourite Tech): The children were observed to know the games well and appeared more confident during this session compared to the previous sessions where videogames were not used. The school SLT and TA agreed with this observation.<br><br>***My dream job is Minecraft You Tuber*** (Child 3, Session 6 - Favourite Tech)<br><br>Field notes (Session 7 - App Design): The children continued the high level of engagement seen in Session 6, they focused well and were motivated to be 'app designers'.<br><br>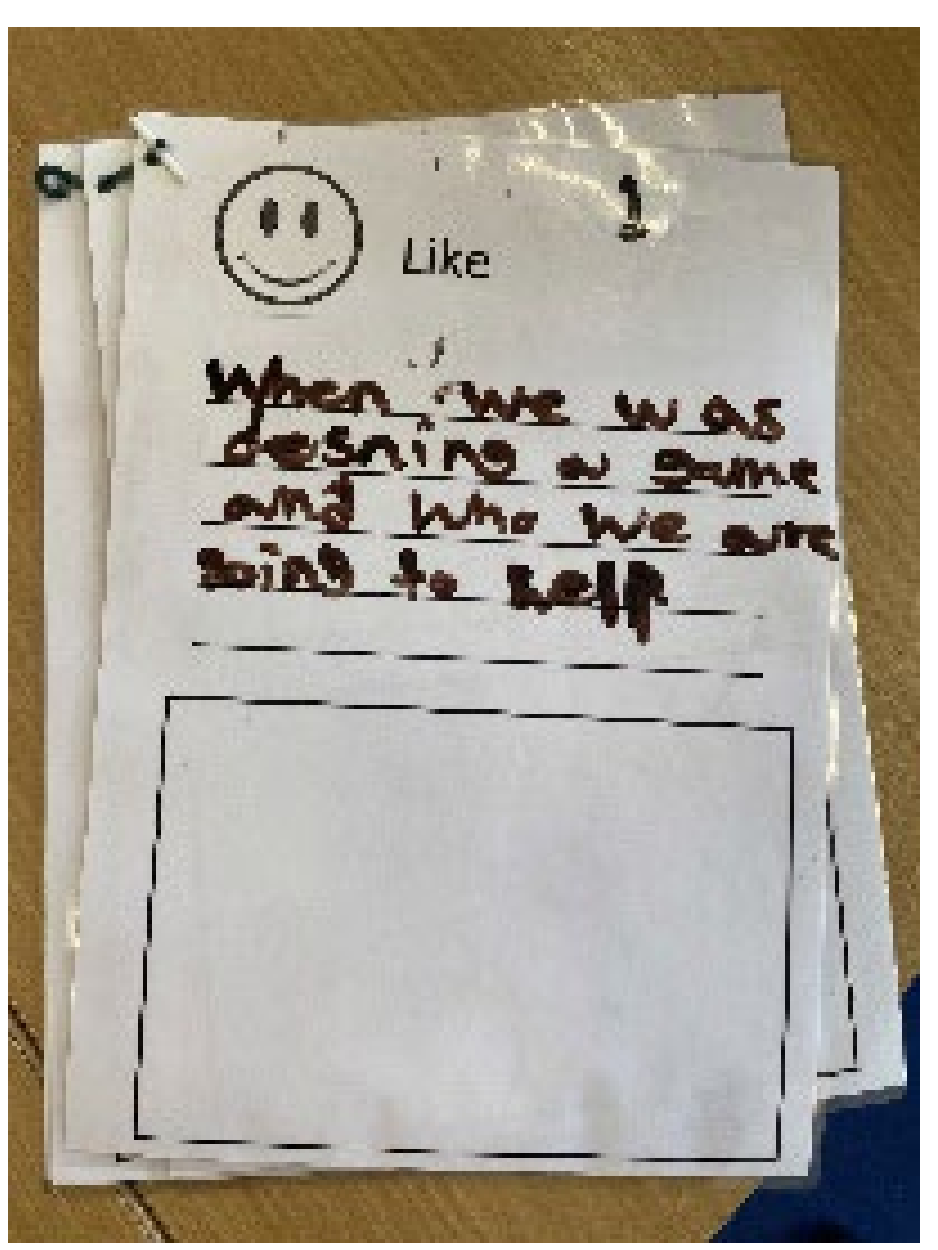<br>**[I liked] *when we was designing a game and who we are going to help*.**<br>(Child 1, Session 7 – App Design) |
| | | Adult support needed to regulate after the videogame immersion session. | Fieldnotes (Session 6 - Favourite Tech): Child 2 became very animated in session when playing the videogames. For example, struggling to stay seated, rocking on chair when seated, hand flapping. Whilst he had needed some verbal TA support to focus in previous session, the TA had to use both verbal prompts and give Child 2 a sensory squeeze toy as well as some 'quiet time' before he was ready to return to class and his next lesson. Also, this was the only session where more children selected tired as their emotion post-session than pre-session.<br><br>TA Comment*: It's great he (Child 2) enjoyed the session but that's overexcitement. In class that would be tricky, he'd need help to regulate* (Session 6 - Favourite Tech) |
| | 4. For a positive experience, the processing demands of design activities need to align with children's capability and capacity | Group activities can be overwhelming but multimodal communication opportunities help | Fieldnotes (Session 2 – Word Games): Child 4 did not speak at all during group activities but did contribute to written tasks. Child 1 and 2 spoke ++ and needed regular prompting from school SLT and TA. Different strategies were explored with both staff and use of sticky notes as an extra communication tool was agreed on. Benefits – removes time pressure, removes demand of speaking in a group, can add to over the session, can be used in addition to and/or instead of other communication methods (choice).<br><br>School SLT comment: *He (Child 4) is much better 1:1. He just gets overwhelmed in a group and tends to withdraw. This happens in class too, but we are working on this in his SLT sessions, and he is getting there* (Session 2 - Word Games).<br><br>Example of sticky notes response when exploring the term *strengths, needs, support* (Child 4, Session 4 – Support Strategies) |

<table>
<tr>
<td></td>
<td></td>
<td></td>
<td>Fieldnotes (Session 6 – Evaluating Strategies): Children able to choose/self-generate a range of support strategies but ++ struggled to complete sentences on what they liked/disliked/would change about a strategy. They either opted out or stated, ‘I don’t know’. School SLT felt this was linked to the verbal reasoning demands of the activity and it was agreed to avoid verbal evaluation tasks in future sessions.<br><br>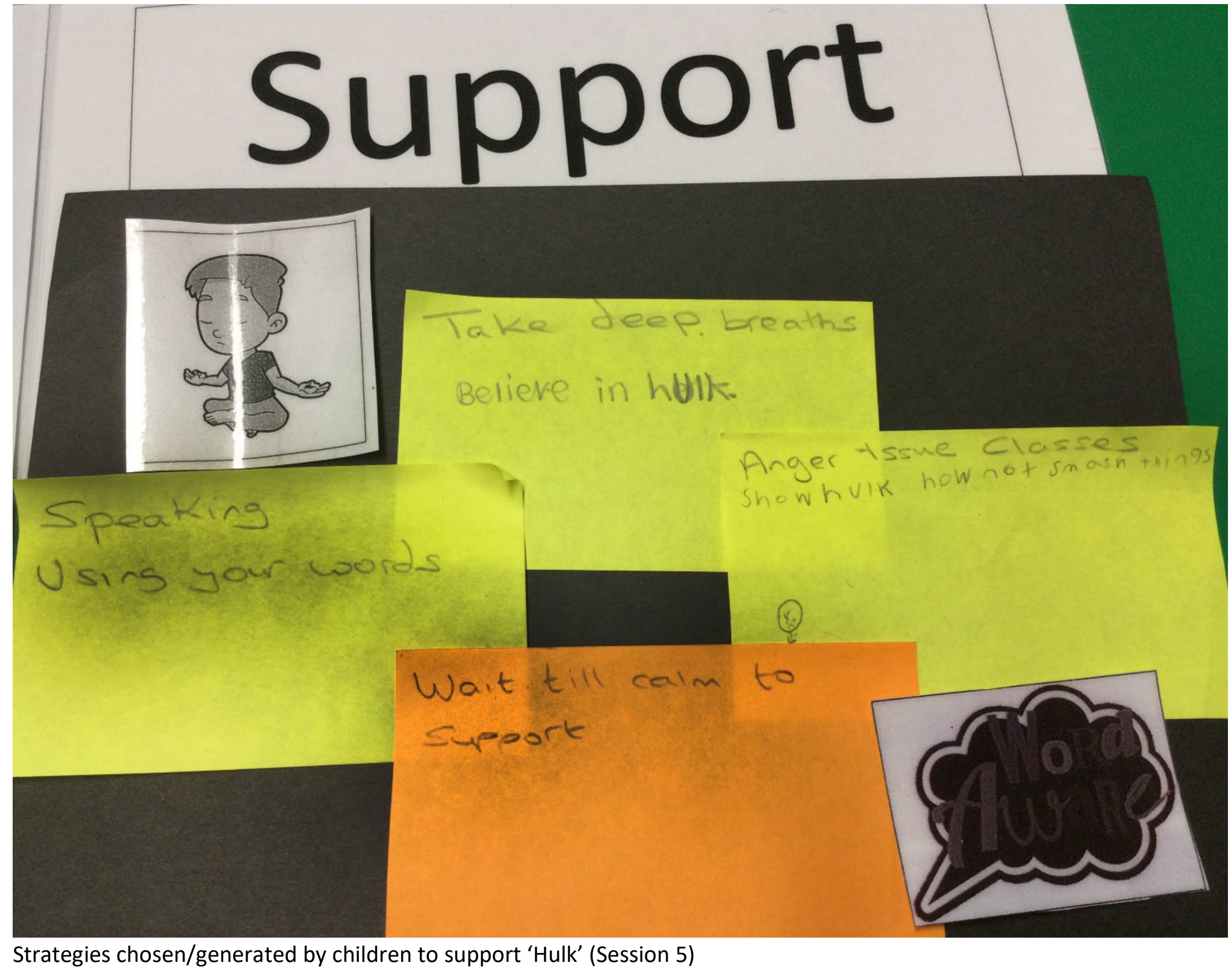

<br>Strategies chosen/generated by children to support ‘Hulk’ (Session 5)</td>
</tr>
</table>

| | | | |
|---|---|---|---|
| | | | School SLT comment: *Coming up with support strategies is easier for them. If you think of Blank levels, there's less reasoning in saying what Hulk could do but the why is harder. We are working on Blank levels in SLT and they are mostly fine with 'what' questions but need scaffolding with the 'why' questions* (Session 6 - Evaluating Strategies)<br>Reference: Blank, M. (2002). Classroom discourse: A key to literacy. In K. Butler & E. Silliman (Eds.), Speaking, reading and writing in children with learning disabilities: New paradigms in research and practice (pp. 151-173). Malwah, NJ: Erlbaum. |
| RQ2: What are the system requirements when designing a word learning tool for neurodiverse children? | 1: Children are responsive to multimedia opportunities for gaining and sharing information but struggle with abstract content | Children use and respond to multiple communication methods depending on personal preference and activity type | Written and spoken:<br>Post-session feedback (Child 1, Session 2 - Word Games)<br>Written and drawn:<br>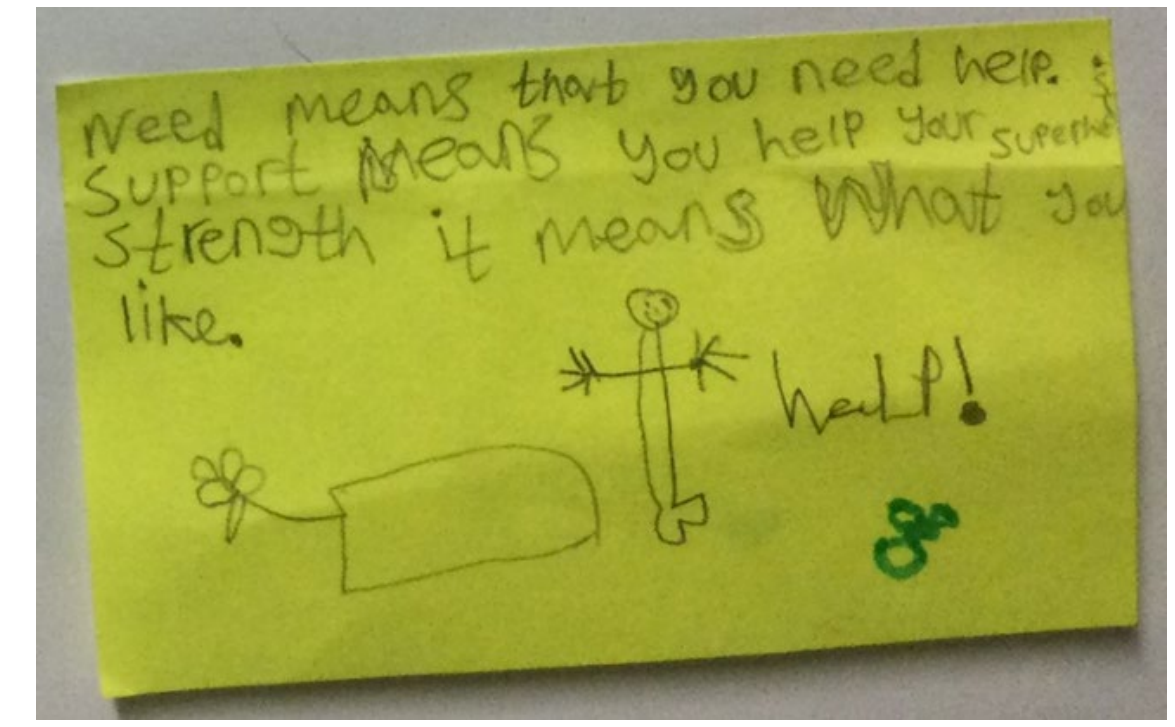<br><br>Word description activity (Child 4, Session 4 - Support Strategies)<br>Role play<br>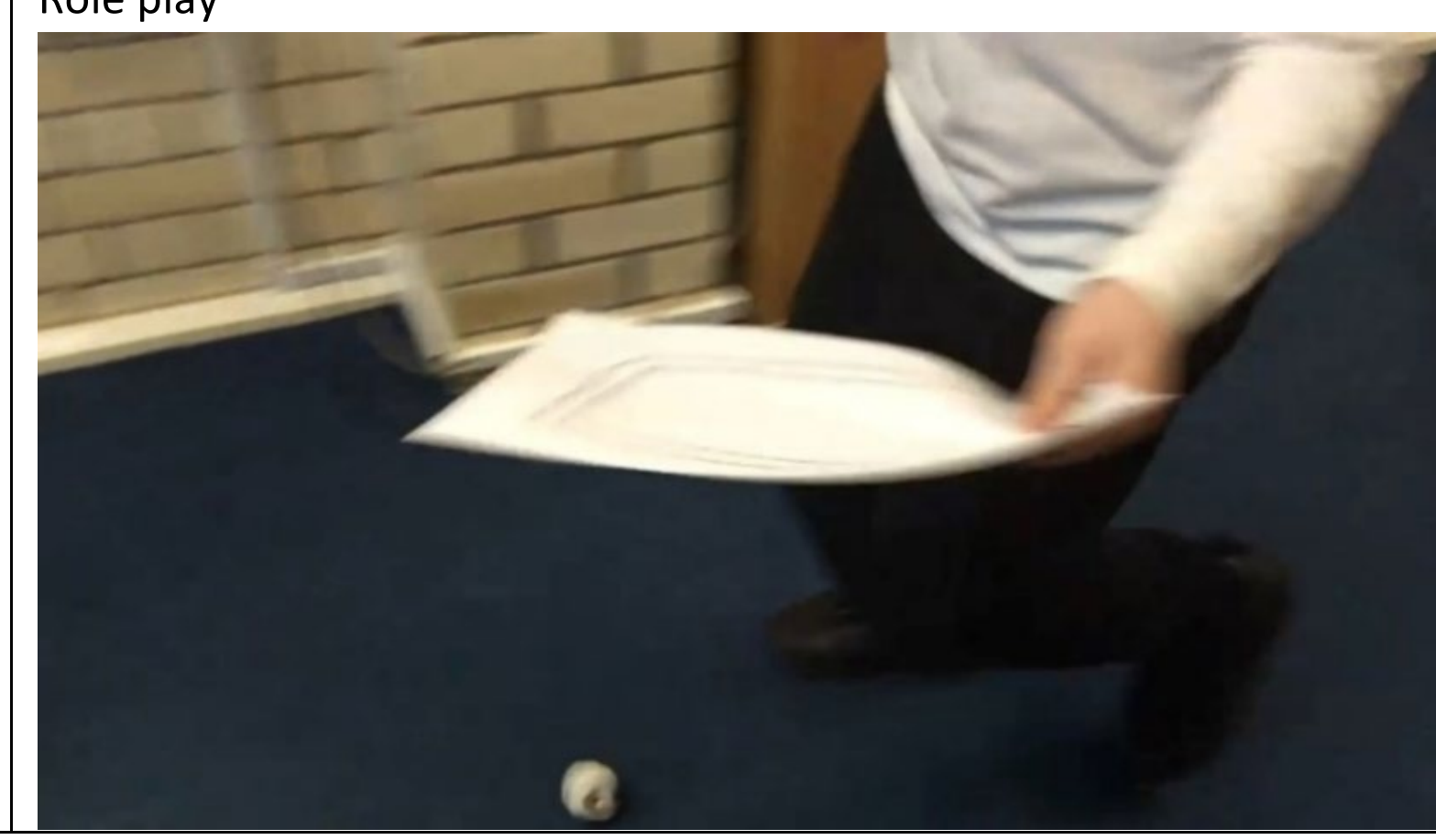<br>App design ideas (Child 4, Session 7 - Design an App) |
| | | Better response to relatable/concrete multimodal stimuli than obscure/abstract media. | 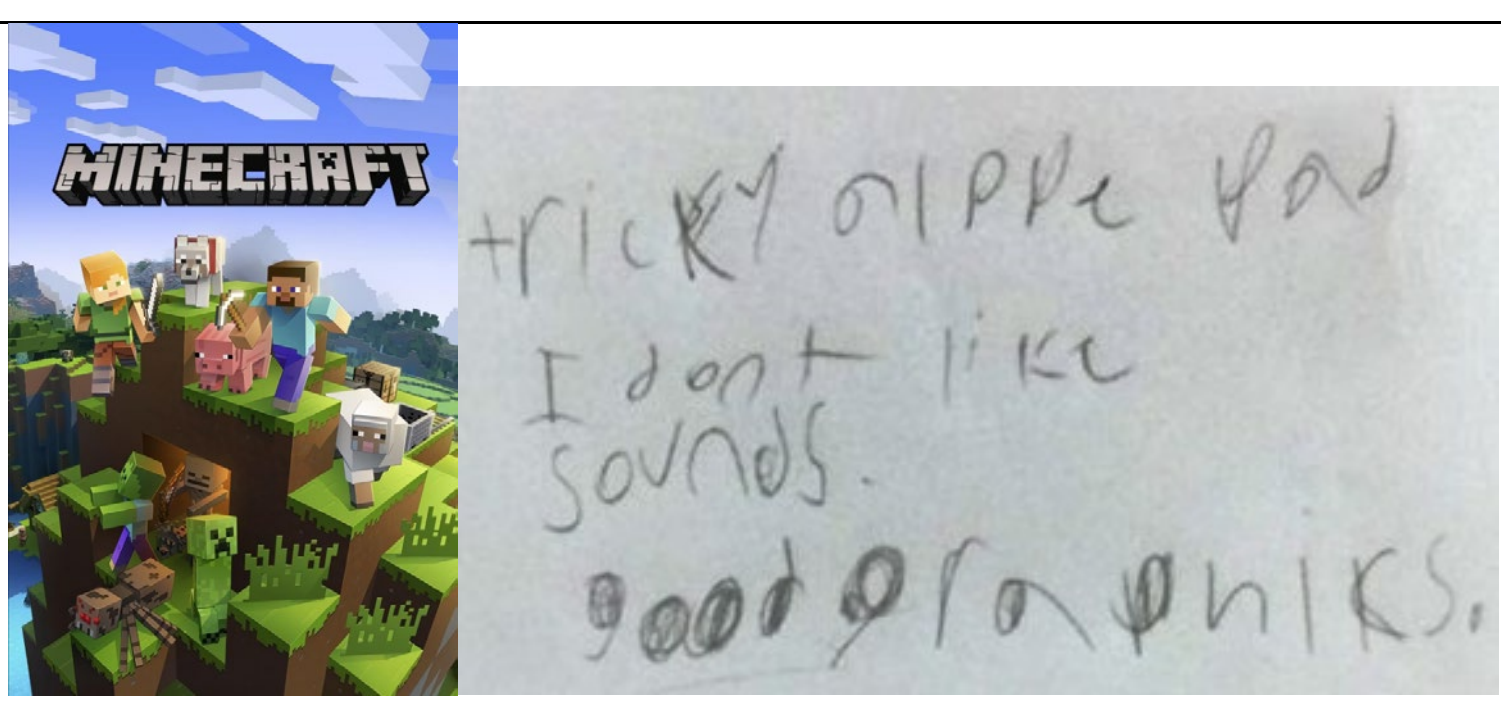<br><br>Written feedback:<br>*Tricky [text illegible]*<br>*I don’t like sounds.*<br>*Good graphics*<br><br>Verbal feedback: ***I don’t like the creepy sounds. They don’t make sense.*** (Child 2, Session 6, Favourite Tech)<br><br>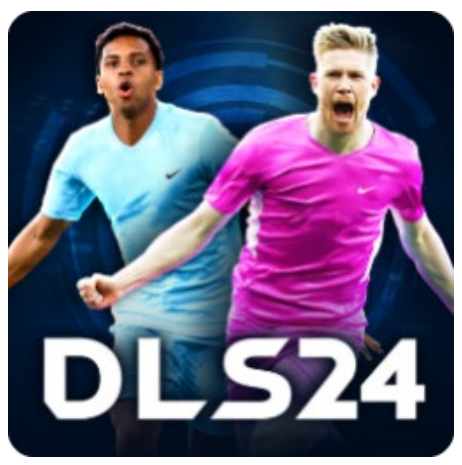<br><br>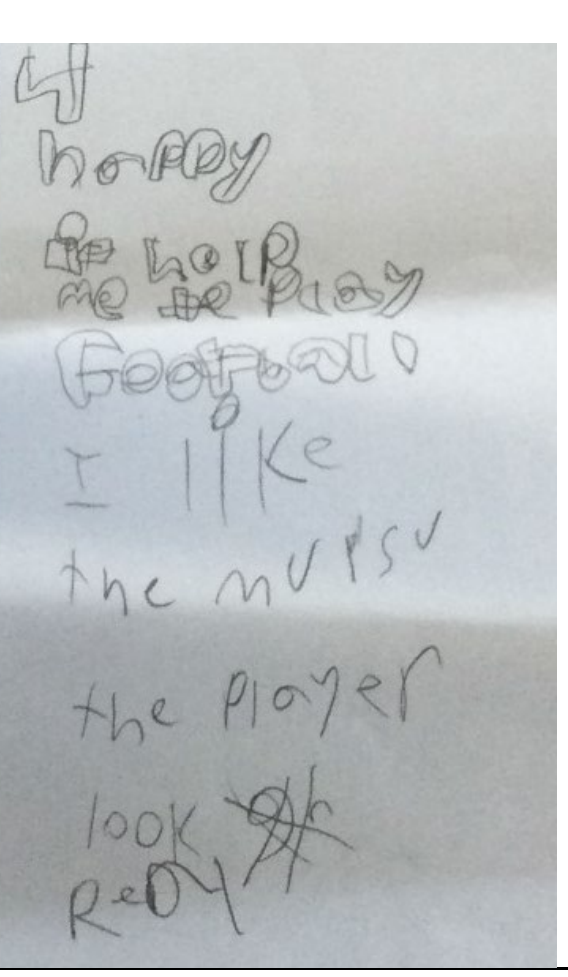<br>Written feedback:<br>*4* (participant number)<br>*Happy*<br>*It help me to play football*<br>*I like the music* (pop songs)<br>*The player look real*<br>(Child 4, Session 6, Favourite Tech)<br><br>Verbal feedback:<br>***Loved it. The players look real. So sick.*** (Child 1, Session 6, Favourite Tech) |
| | 2. Customising learning to children's interests encourages a positive attitude | Children relate strengths to things they are interested in. | Exploring key words – summary of children defining the word 'strength' (Session 3 – Strengths & Needs)<br><br>Detailed verbal responses:<br>***My strengths are my favourite things, things I good at.*** (Child 1, Session 3 – Strengths & Needs)<br>***Your strengths are things you like.*** (Child 2, Session 3 – Strengths & Needs) |
| | | Greater interest in customised activities. | Fieldnotes (Session 6 – Favourite Tech): Both videogames allow a high degree of individualisation. In Minecraft children craft their own digital world, selecting materials, types of structures, colours, and layout. DLS24 enables users to create their 'dream team' by selecting players and personalising kits, logos, and stadiums of characters. This tailoring to individual preferences seemed to appeal.<br><br>***I like you can customise it* [Feedback on DLS24]** (Child 1, Session 6 - Favourite Tech)<br><br>***Look what I made! It's mine!* [Feedback on Minecraft]** (Child 3, Session 6 - Favourite Tech) |
| | 3. Repeated, child-led learning opportunities promote skills | | ***The more you do, the better you get*** (Child 3, Session 4 – Support Strategies)<br><br>***In the first screen we need a pop-up with rules, so we know every time what to do*** (Child 2, Session 7 – App Design) |
| | 4: Learning through technology must be embedded in the | Children value adult support during word learning | ***I would ask teacher for help*** (Child 3, Session 4 - Support Strategies)<br><br>***You can ask Google, but you can't ask Google always. You need to ask people. People know things*** (Child 2, Session 4 - Support Strategies) |

|  | real world to complement existing practice and align with children's individual circumstances | Children included word-learning support strategies relating to their general state (readiness for learning) | ***Wait till they're ready to liste****n* (Child 4, Session 4 - Support Strategies)<br><br>***Taking deep breathes helps*** (Child 1, Session 4 - Support Strategies) |
|---|---|---|---|
|  |  | Children find linking the real and digital world engaging (augmented reality, mixing real media and cartoon style) | Child 4, Session 7 – App Design.<br><br>Setup: Child 4 drew an iPad on a sheet of paper, then used the iPad to 'record' himself playing football with another child (using a mini football prop for a choice of everyday item props). Principal researcher asked: *Tell me how the iPad is being used:*<br>*The ipad is being there with a video cam on it…and then you look…and then do you know that thing called Pokémon Go where you use like that (demonstrates holding up paper with iPad drawn on it and directing at mini football prop)…and then you see a football… and then you see a person (moves 'paper iPad' from football to another child) …and then you get the ball?*<br>*Yes, Pokémon Go* (Principal researcher)<br>*Right, it's like Pokémon Go.* ***So, there should be cartoons and then real things, like for Pokémon Go****, but you don't get Pokémon you try to tackle.* |
|  | 5. Playfulness and error-less learning are key motivators | Open-ended exploration is appealing | Fieldnotes (Session 6 - Favourite Tech): Focus in Minecraft is exploration and creativity which appealed to one child in particular. In contrast, DLS24 is focused on winning a football game. This was highly motivating to some of the children, however, as one child was not interested in football they did not engage as much.<br><br>*I like it cos there's* **no such thing as game over** [Minecraft Feedback] (Child 2, Session 6 - Favourite Tech) |

| | | Indicators to track progress are appealing | ***There should be levels, like in video games*** (Child 3, Session 7 – App Design).<br>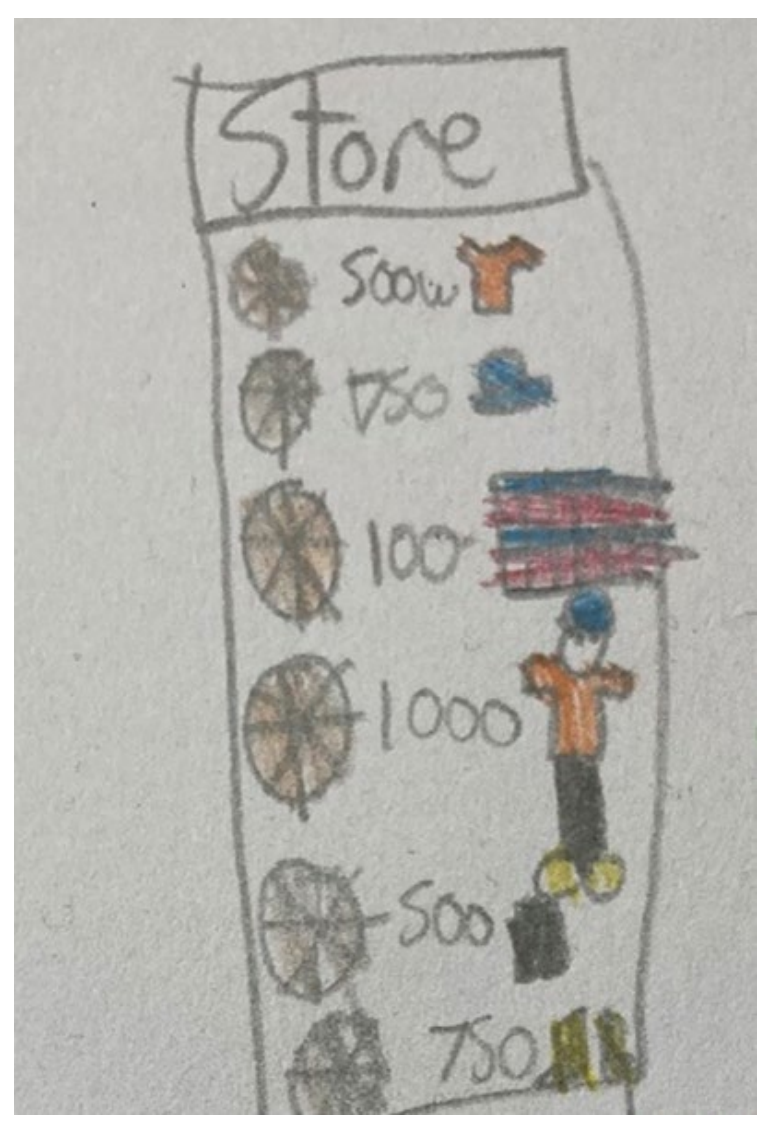<br><br>Drawing of a reward system for the word-learning app (Child 3, Session 7 – App Design)..<br>Verbal description: *So, say you were working on clothes, each time you finish a bit something changes. Like you get more points, but the points are for doing more bits, and* ***game hardness is the same*** (Child 1, Session 7 – App Design)<br>***Yeah, that makes sense…when you do more learning, you fill more of the bar, and this shows how well you're doing.*** (Child 2, Session 7 – App Design). |
|---|---|---|---|